\documentclass[twocolumn,superscriptaddress,secnumarabic,aps,prb,nofootinbib,numerical]{revtex4}
\usepackage[english]{babel}
\usepackage{epstopdf}
\usepackage{pdfpages}
\usepackage{float}
\usepackage{enumerate}

\usepackage{amsmath} 
\usepackage{dsfont}
\usepackage{cancel}
\usepackage{amsthm} 
\usepackage{amssymb}	
\usepackage{graphicx} 
\usepackage{empheq}
\usepackage[absolute,overlay]{textpos}
\usepackage{xcolor}

\allowdisplaybreaks

\let\baraccent=\= 
\renewcommand{\=}[1]{\stackrel{#1}{=}} 

\DeclareMathOperator{\Tr}{Tr}

\begin{document}

\title{{\color{blue} Edge mode manipulation through commensurate multifrequency driving}}

\date{\today}

\author{Paolo Molignini}
\affiliation{Clarendon Laboratory, Department of Physics, University of Oxford, OX1 3PU, United Kingdom}







\begin{abstract}
We explore the impact of commensurate multifrequency driving protocols on the stability of topological edge modes in topological 1D systems of spinless fermions. 
Using Floquet theory, we show that all the topological phase transitions can be mapped to their single-frequency counterparts in terms of effective driving parameters.
This drastic simplification can be explained by considering the gap closures in the quasienergy dispersion.
These gap closures are pinned to high-symmetry points, where all the effective Floquet Hamiltonians collapse to the same form.
While for all protocols all topological phase transitions coincide, the gap size and the number of edge modes in the quasienergy spectra vary considerably depending on the chosen driving pattern.
This gives rise to a full range of edge states with different degrees of localization, from highly localized to completely delocalized.
Switching between different driving protocols then suggests a dynamical control of the localization and stability of topological edge modes, with possible applications in quantum computation.
We illustrate our findings on three paradigmatic fermionic systems --- namely the Kitaev chain, the Su-Schrieffer-Heeger model, and the Creutz ladder, and demonstrate how to control the localization length of the corresponding edge states.

\end{abstract}

\maketitle

\section{Introduction}
The discovery of topological order has drastically altered our understanding of phase transitions~\cite{Landau,Miransky-book,Wen:1990,Thouless:1982, Wen:1989, Qi:2008}.
Topological states represent radically new phases of matter with robust ground state degeneracy and complex highly-entangled structures that lead to outlandish new properties~\cite{Goldman:1995, dePicciotto:1997, Martin:2004, Moore, Kitaev:2001, Mong:2014, Wen:1995, XieChen:2010,Fidkowski:2010}. 
Arguably, the most paradigmatic of such properties is the emergence of perturbatively-stable modes at the edges of a finite-size sample.
Depending on the system, these modes can be fermionic or anyonic, abelian or non-abelian, static or dispersive, and therefore can be harnessed for countless different technological applications ranging from quantum computation~\cite{Kitaev:2003, Nayak:2008}, to low-power electronics~\cite{Vandenberghe:2017} and spintronics~\cite{WangBook:2016}.
 
A particularly interesting class of topological order arises in periodically driven or Floquet systems, to which
belong \textit{e.g.} Floquet topological insulators~\cite{Kitagawa:2010,Lindner:2011, Cayssol:2013, Harper:2017, Roy:2017,Esin:2018}, Floquet topological superconductors~\cite{Liu,Thakurathi:2013, Benito:2014,Thakurathi:2014,Wang:2014,Sacramento:2015,Thakurathi:2017,Molignini:2017,Molignini:2018,Cadez:2019}, and Floquet  semimetals~\cite{Wang:2014-Weyl-semimetal, Chan:2016-Weyl, Yan:2016, Bucciantini:2017, Huebener:2017, Cao:2017, ChenZhou:2018, Li:2018}.
For these systems, tuning energy parameters periodically in time can realize edge modes in the quasienergy spectrum even in parameter ranges where the static counterparts are topologically trivial.
Furthermore, because of the time-periodicity, the energy bands are backfolded to a Floquet-Brillouin zone at which boundaries a new flavor of edge modes --- so-called $\pi$ edge modes --- can appear.
The interplay between 0 and $\pi$ edge modes gives rise to very rich phase diagrams with a tunable cascade of topological phase transitions (TPTs).

A natural question emerging in periodically driven topological systems is how the topological features generalize when the system is driven not with just one frequency, but several.
In recent years, numerous studies have been conducted in this direction~\cite{Martin:2017,Peng:2018,Peng2:2018,Crowley:2019,Crowley2:2019}.
The literature has however focused mainly on incommensurate multifrequency driving, where there is no global periodicity and hence no usual stroboscopic Floquet operator can be defined.
The interplay between different incommensurate drives is instead studied by operating in a multidimensional Fourier space, which effectively induces synthetic lattice dimensions.
The synthetic dimensions can be explored to induce energy conversion between different driving modes~\cite{Martin:2017,Crowley2:2019}, multiplex edge modes in lower dimensional geometries~\cite{Peng:2018, Peng2:2018}, or stabilize dynamical topological phases~\cite{Crowley:2019}.
However, while the study of incommensurate multifrequency driving has opened intriguing avenues for novel quantum engineered topological states of matter, the effect of \textit{commensurate} multifrequency driving has been less researched.

In this work, we explore the effect of commensurate multifrequency driving on 1D systems of spinless fermions hosting topological phases (class AIII, D, and BDI in the Altland-Zirnbauer classification of topological systems~\cite{Altland:1997,Schnyder:2008, ChiuReview:2016}).
By employing Floquet theory, we will demonstrate that the introduction of commensurate multifrequency driving has exciting repercussions on the properties of the edge states.
While our discussion is valid for general $2\times 2$-Hamiltonian in $k$-space, we will show concrete examples on three paradigmatic topological models: the Kitaev chain, the Su-Schrieffer-Heeger model, and the Creutz ladder.
By comparing the phase diagrams of the single and multi-frequency driving, we discover that all the TPTs generated from multifrequency driving can be mapped to the single driving TPTs by an appropriate choice of effective driving parameters.
We characterize this mapping analytically in terms of gap closures in the quasienergy dispersion with the help of the bulk-edge correspondence.
Furthermore, we determine that it is possible to dynamically alter the generation and stability of edge modes by carefully choosing an appropriate multifrequency driving protocol.
We show that dynamically switching between different multifrequency driving configurations provides an efficient way to manipulate the quasienergy gap and hence the localization length of the edge modes.
We find that the localization of the edge modes can be controlled over a very wide range, from strong localization to complete delocalization. 
The control of the localization of the edge modes, specially non-abelian Majorana modes, is a key ingredient for possible applications in topological quantum computing~\cite{Nayak:2008, Marra:2017,  Bauer:2018, Stenger:2018, Haim:2019}.
Our results provide a fresh approach to this task by carefully exploiting the versatility of Floquet engineering.

This article is structured as follows.
In section~\ref{sec:multifreq}, we summarize the methods of Floquet theory that we use to analyze the topology of the multifrequency-driven systems and provide an analytical explanation of our findings by analyzing the gap closures of the quasienergy dispersion.
In the following sections, we demonstrate the generality of our approach by applying it to paradigmatic systems belonging to three different symmetry classes: the Kitaev chain in section~\ref{sec:Kitaev}, the Su-Schrieffer-Heeger model in section~\ref{sec:SSH}, and the Creutz ladder in section \ref{sec:Creutz}.
A conclusion in section \ref{sec:conclusions} wraps up our discussion and provides an outlook to future investigations.

\section{Methods}
\label{sec:multifreq}

We consider a time-dependent Hamiltonian $\mathcal{H}(t)$ describing a topological 1D system of spinless free fermions and subjected to periodic driving with multiple \textit{commensurate} periods $T_1$, $T_2$, etc. that fulfill the commensuration condition
\begin{equation}
p_i T_i = p_j T_j = T, \forall i,j
\label{commensuration-rule}
\end{equation}
with $T$ the least common multiple of all the periods.
This kind of multifrequency driving generates complex structures within the full period $T$.
The simple case of single-frequency drive is recollected setting $p_i=p_j, \forall i,j$.

To characterize the topology emerging in the driven multifrequency case, we examine the quasienergy spectrum $\{ \tilde{\epsilon}_{\alpha} \equiv \epsilon_{\alpha} T \}$ of the open chain (normalized between $[-\pi, \pi)$). 
The quasienergies are the eigenvalues of the effective stroboscopic Floquet Hamiltonian
\begin{equation}
h_{\text{eff}} \equiv i \log U(T,0),
\end{equation}
where
\begin{equation}
U(T) \equiv \mathbb{T} \left[ \exp \left( -i \int_0^T \mathrm{d}t \: \mathcal{H}(t) \right) \right]
\label{eq:Floquet-operator}
\end{equation}
is the Floquet operator, \textit{i.e.} the time-evolution operator over one cycle of the drive.
The symbol $\mathbb{T}$ indicates time-ordering.
For simplicity, we have set $\hbar=1$ and the initial instant of the time evolution at $t=0$.
Because of the Floquet theorem~\cite{Shirley}, the quasienergies are $2\pi$-periodic and can therefore be defined in the interval $[-\pi,\pi]$, often termed Floquet-Brillouin zone.
Besides the usual gap closures and associated edge modes at $\epsilon=0$, this backfolding procedure creates new channels for band gap closures at $\pm \pi$, where new kinds of edge states, called $\pi$-modes, can appear~\cite{Kitagawa:2010,Lindner:2011}.
The number of gapped $0$- and $\pi$-quasienergies determines the total number of localized edge states~\cite{Thakurathi:2013}.

We remark that the description with the effective Floquet Hamiltonian might start to diverge from the real instantaneous state occupation when the driving frequency is low enough to allow quasidegenerate eigenstates of the effective Hamiltonian to interact strongly and hybridize, thereby opening up heating channels. 
Nevertheless, many recent theoretical~\cite{Dalessio:2014,Abanin:2015,Bukov:2016,Mori:2016,Kuwahara:2016,Weinberg:2017,Abanin:2017,Abanin2:2017} and experimental~\cite{Jotzu:2014,Messer:2018,Cheng:2019,Rubio-Abadal:2020} studies have demonstrated that in the high-frequency (low period) regime heating can be exponentially suppressed and the effective Floquet Hamiltonian is a valid description.
While these studies typically deal with harmonic drives, they can be extended to other drives by considering a Fourier decomposition and by taking the highest harmonic in the approximation as the relevant frequency that might lead to resonance.
Other studies have also separately shown the validity of the effective Floquet Hamiltonian for periodically kicked systems, such as the ones we analyze in this article~\cite{Vajna:2018}.
Topological many-body states in periodically driven systems, although generally metastable, can therefore have very long lifetimes~\cite{Abanin:2015}.
Coupled with their strong energy localization, this should allow stroboscopic edge modes to be populated over time scales relevant enough for dynamical manipulations.

\subsection{Connection between single and multifrequency driving}

To locate the TPTs where the number of edge modes changes, and therefore map out the topological phase diagram, we numerically count the number of quasienergies below a given threshold $\delta$ around $\epsilon=0$ and $\epsilon=\pi$. 
In general, multifrequency driving protocols with different commensurations will lead to different quasienergy spectra because of the different product of exponentials making up the Floquet operator.
However, we will now show that, under appropriate conditions, the gap closures and hence the topological phase boundaries can always be mapped to the single-frequency case.

To prove this, we exploit the bulk-edge correspondence~\cite{Bernevig13,ProdanBook:2016} to extract information about the topology from the bulk Floquet Hamiltonian defined for the infinite system, or the system with periodic boundary conditions, where the momentum $k$ is a good quantum number.
For the generic 1D spinless fermionic systems analyzed here, the static bulk Hamiltonian is a $2\times2$ complex matrix that can be parametrized as~\cite{LinhuLi:2016}
\begin{equation}
\mathcal{H}(k, \vec{M}) = h_x(k, \vec{M}) \sigma^x + h_y(k, \vec{M}) \sigma^y + h_z(k, \vec{M}) \sigma^z,
\end{equation}
where $\vec{M}$ is a vector collecting all possible energy parameters that can be tuned in time.
Note that this general form can encompass also systems with longer-range hoppings or pairings that do not necessarily have gap closures at inversion symmetric points $k_0=0,\pi$.

The presence of symmetries will further reduce the form of the Hamiltonian.
Classes AIII and BDI obey a chiral symmetry $\mathcal{S}$, \textit{e.g.} $\mathcal{S}=\sigma^y$ without loss of generality~\footnote{The other symmetry operators $\sigma^x$ and $\sigma^z$ can be obtained by unitary rotations of the Hamiltonian, which do not change its physics.}, implying $h_y \equiv 0$.
Note that in the BDI, time-reversal and particle-hole symmetries will further reduce the form of the Hamiltonian, but this is not necessary for the results we want to obtain.
For the remaining topologically nontrivial class of spinless fermions in 1D, namely class D, only charge conjugation $\mathcal{C}$ exists.
Again without loss of generality, we can write $\mathcal{C} = \mathcal{K} \circ \sigma^x$, where $\mathcal{K}$ is complex conjugation.
This symmetry then restricts the form of the coefficients for the Pauli matrices to be $h_x(k)=-h_x(-k)$, $h_y(k)=-h_y(-k)$, $h_z(k)=h_z(-k)$.

Now, let us assume we drive in time a given parameter: $\vec{M} \ni M_d \to M_d(t) = M_{d,1}(t) + M_{d,2}(t) + \cdots $, where $M_{d,1}(t)$ has $T_1$ periodicity, $M_{d,2}(t)$ has $T_2$ periodicity etc. 
As a concrete example, we can take $M_{d,i}(t)= M_{d,i} T_i \sum_{n \in \mathbb{Z}} \delta( t - n T_i)$, but other periodic drives lead to similar results (see \textit{e.g.} appendix \ref{app:harm-KC}).
If there is a high-symmetry point (HSP) $k_0$ for which
\begin{equation}
h_x(k_0, \vec{M})=0, \quad \forall \vec{M},
\label{eq:condition-mapping}
\end{equation}
the TPTs for a multifrequency driving protocol can be mapped back to the ones in the single driving protocol in terms of an effective driving intensity $M_{\text{eff}} = M_{d,1} + M_{d,2} + \cdots$.
Note that condition~\eqref{eq:condition-mapping} is rather broad, because in lattice models $h_x$ is typically given as a trigonometric polynomial in $k$, for which we can always find a root for all parameters.

To understand this mechanism, let us again consider the bulk Floquet operator, which for the generic 1D fermionic systems analyzed here takes the following form constraint by unitarity:
\begin{equation}
U_k(T) = \left(
\begin{array}{cc}
A & B \\
-B^* & A^*
\end{array}
\right).
\end{equation}
\label{Floquet-operator-momentum}
The eigenvalues of the $2\times2$ matrix $U_k(T)$ can be written in a compact form as
$
\lambda_k^{\pm} = \frac{\Tr U_k(T)}{2} \pm \sqrt{ \left(\frac{\Tr U_k(T)}{2} \right)^2 - \det U_k(T)}.
$
Knowing additionally that $\det U_k(T) =1$, we can use the identity $\arccos z = -i \log \left( z + \sqrt{z^2 -1} \right)$ to extract the effective Floquet quasienergy dispersion $\theta_k$, defined via $U_k(T) \psi(k) = e^{i \theta_k} \psi(k)$, as
\begin{equation}
\theta_k = -i \log \lambda_k^+ 
= \arccos \left[ \Re[A] \right].
\end{equation}
%
The gap closures of the Floquet quasienergy dispersion will determine the TPTs.
For classes AIII and BDI, condition \eqref{eq:condition-mapping} implies that the bulk Hamiltonian at the HSP is simply $\mathcal{H}(k_0, \vec{M}) = h_z(k_0, \vec{M}) \sigma^z$.
For class D, only the HSPs $k_0=0,\pi$ are allowed\footnote{This follows from $h_x$ and $h_y$ being odd functions in $k$, which can be decomposed as sums of sine functions which are zero only at $k_0=0,\pi$.}, which also yields $\mathcal{H}(k_0, \vec{M}) = h_z(k, \vec{M}) \sigma^z$.
Therefore, the matrix exponentials appearing in $U_{k_0}(T)$ will all commute with each other, and we can simply sum up the different intensities in the exponents into a single effective parameter $M_{\text{eff}} \equiv M_{d,1} + M_{d,2} + \cdots$.
In terms of $M_{\text{eff}}$, the Floquet quasienergy dispersion will have the same form as in the single driving case, and the TPTs will then coincide.

Note that this analysis only focused on \textit{spinless} fermionic models. 
In 1D, two other classes are topologically nontrivial, namely class DIII and CII. 
However, because for these classes time reversal symmetry fulfills $\mathcal{T}^2=-1$, the systems must  describe spinful fermions, with a $4\times4$ bulk Hamiltonian parametrized as $H(k) = \sum_{i,j=0,x,y,z} h_{i,j} s_i \otimes \sigma_j$, where $s_i$ ($\sigma_j$) are Pauli matrices in spin (particle-hole) space.
Because this form is much more involved, a possible analysis of the reduction from multifrequency to single frequency driving in such classes is left for future studies.

We will now apply our general analysis above to concrete systems belonging to all three different symmetry classes.

\section{Kitaev chain}
\label{sec:Kitaev}

We begin illustrating our results by considering the Kitaev chain (KC) subjected to various kinds of multifrequency driving schemes. 
The KC has become a paradigmatic example of a topological superconductor hosting Majorana (\textit{i.e.} anyonic) edge modes~\cite{Kitaev:2001}.
Numerous proposals, for instance with proximitized semiconductor nanowires~\cite{Mourik:2012}, magnetic adatoms~\cite{Nadj-Perge:2014}, or arrays of quantum dots~\cite{Sau:2012}, have sought to realize it experimentally with preliminary observable signatures of their physical existence.

The real-space Hamiltonian of the KC is
\begin{align}
\label{Hamiltonian_fermionic}
\mathcal{H}_{KC} &= \sum_{n=1}^{N-1} \bigg[ \tau f^{\dagger}_n f_{n+1}  + \Delta f_n f_{n+1} + h.c. \bigg] \nonumber \\
& \quad - \sum_{n=1}^{N} \mu_0 (2 f^{\dagger}_n f_n -1)
\end{align}
and describes a one-dimensional (1D) chain of $N$ spinless fermions with operators $f^{(\dagger)}_n$ fulfilling anticommutation relations $\left\{ f^{\dagger}_m, f_n \right\} = \delta_{mn}$, $\left\{ f_m, f_n \right\} = \left\{ f^{\dagger}_m, f^{\dagger}_n \right\} = 0$.
The particles are coupled together through a nearest-neighbor hopping term $\tau$, a $p$-wave superconducting pairing $\Delta$, and a chemical potential $\mu_0$.

The KC with open boundary conditions hosts zero-energy states --- dubbed ``Majorana zero-energy modes" (MZMs) --- whose eigenvectors are purely real and localized at the ends of the chain~\cite{Kitaev:2001, Thakurathi:2013}.
Albeit the end modes can be detected only for long open chains, the bulk-edge correspondence~\cite{Bernevig13, ProdanBook:2016} allows us to characterize them through a topological invariant constructed in momentum space, \textit{i.e.} for a system with PBC $f^{(\dagger)}_{N+1}=f^{(\dagger)}_1$.
Introducing the Fourier transform $f_k = \frac{1}{\sqrt{N}} \sum_{n=1}^N f_n e^{ikn}$ (and similarly for $f_k^{\dagger}$) the Hamiltonian transformed to momentum-space reads
\begin{equation}
\mathcal{H}_{KC} =  \sum_{0 \le k \le \pi}  \left( \begin{array}{cc} f^{\dagger}_k & f_{-k} \end{array} \right) \mathcal{H}_{KC,k}  \left( \begin{array}{c} f_k \\ f^{\dagger}_{-k} \end{array} \right)
\end{equation}
with the Bogoliubov-de Gennes (BdG) bulk Hamiltonian 
\begin{equation}
\label{Hamiltonian_momentum}
\mathcal{H}_{KC,k} = 2 \Delta  \sin k \:  \sigma^y +  2\left(  \tau \cos k - \mu_0 \right) \: \sigma^z,
\end{equation}
where $\sigma^i$ denote Pauli matrices.



\subsection{Multifrequency driving in the Kitaev chain}

One possible driving scheme consists of driving the same parameter (\textit{e.g.} $\mu$) as 
\begin{equation}
\mu(t) = \mu_0 + \mu_1(t) + \mu_2(t),
\end{equation}
where the two functions $\mu_1(t)$ and $\mu_2(t)$ are periodic with different but commensurable periodicities $T_1, T_2$ such that $p_1T_1 + p_2 T_2 =T$, $p_1,p_2 \in \mathbb{N}$.
Alternatively, different parameters could be driven simultaneously, \textit{e.g.} $\tau(t)=\tau_0 + \tau_1(t)$ and $\mu(t)=\mu_0 + \mu_2(t)$, where the drivings should again have different but commensurate periods. 
In terms of their effect on the topology of the system, the two different driving schemes are equivalent, as we can verify by examining the bulk Kitaev Hamiltonian.
It is known in the literature that the topology of the KC is exclusively determined by the $\sigma^z$ component~\cite{LinhuLi:2016,Molignini:2018}.
This is because the gap is locked at the high-symmetry points (HSPs) $k_0 = 0, \pi$ due to inversion symmetry, and there the $\sigma^y$ component is zero.
Therefore, periodic driving of the hopping $\tau(t)$ should be topologically equivalent to driving the chemical potential $\mu(t)$, while driving $\Delta(t)$ should not change the topology at all. 
Note that this reasoning excludes frozen dynamics phenomena and other kinds of anomalous transitions at $k_0 \neq 0,\pi$~\cite{Molignini:2017}, which are however less systematic and therefore will be ignored here.

\begin{figure}
\centering
\includegraphics[width=\columnwidth]{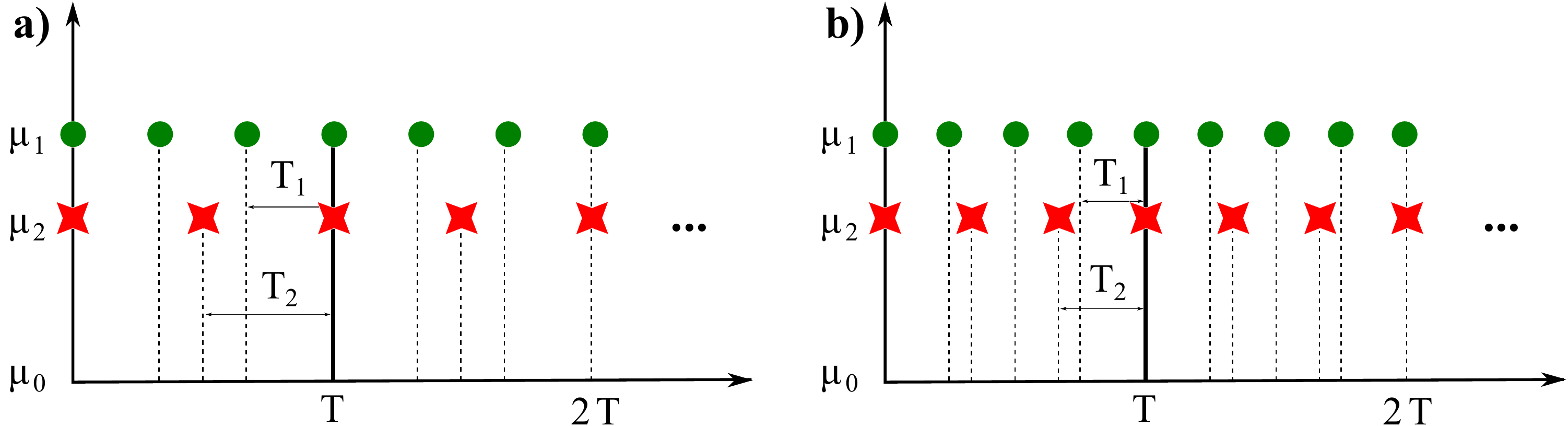}
\caption{Illustration of two possible protocols for Dirac delta multifrequency driving with noninteger ratios: a) 3:2 drive, b) 4:3 drive.}
\label{fig:multifrequency-protocols}
\end{figure}

We will now consider a practical realization of multifrequency driving in terms of Dirac pulses:
\begin{align}
\mu(t) &= \mu_0 + \mu_1 T_1 \sum_{m \in \mathbb{Z}} \delta(t - mT_1) \nonumber \\
& \quad \quad \:\: + \mu_2 T_2 \sum_{m' \in \mathbb{Z}} \delta(t - m' T_2).
\label{mu-double-frequency-Kitaev-delta}
\end{align}
Note that this choice of driving scheme does not impact the generality of our results.
In appendix~\label{app:frozen-SSH}, we show that the mapping to the single-frequency parameter space can also be performed for a harmonic drive.
For each different commensuration, the topological phase diagram can be obtained by diagonalizing the effective Floquet Hamiltonian in the Majorana representation.
Similarly to single-frequency driving, the time ordering in the Floquet operator can be evaluated directly when subjected to a series of delta pulses, and the Floquet operator can be calculated analytically as a product of exponentials.
An exemplary depiction of two multifrequency driving protocols is offered in Fig.~\ref{fig:multifrequency-protocols}.

Much like in the static case, the quasienergy spectrum of the driven case can exhibit Majorana zero-energy modes (MZMs) localized at the ends of the chain.
Because of the periodicity in the quasienergy spectrum, though, it is also possible to generate a new flavor of Majorana modes at quasienergy $\pm \pi$.
We will refer to these as Majorana $\pi$ modes (MPMs).
Much like MZMs, MPMs are purely real and localized at the edges~\citep{Thakurathi:2013}.
However, in contrast to the static case, it is possible to generate a hierarchy of MZMs and MPMs by simply tuning the system's parameters over a wide range.
The system can hence be made topological even when the undriven phase has trivial topology~\citep{Thakurathi:2013, Molignini:2017}.

\begin{figure}
\centering
\includegraphics[width=\columnwidth]{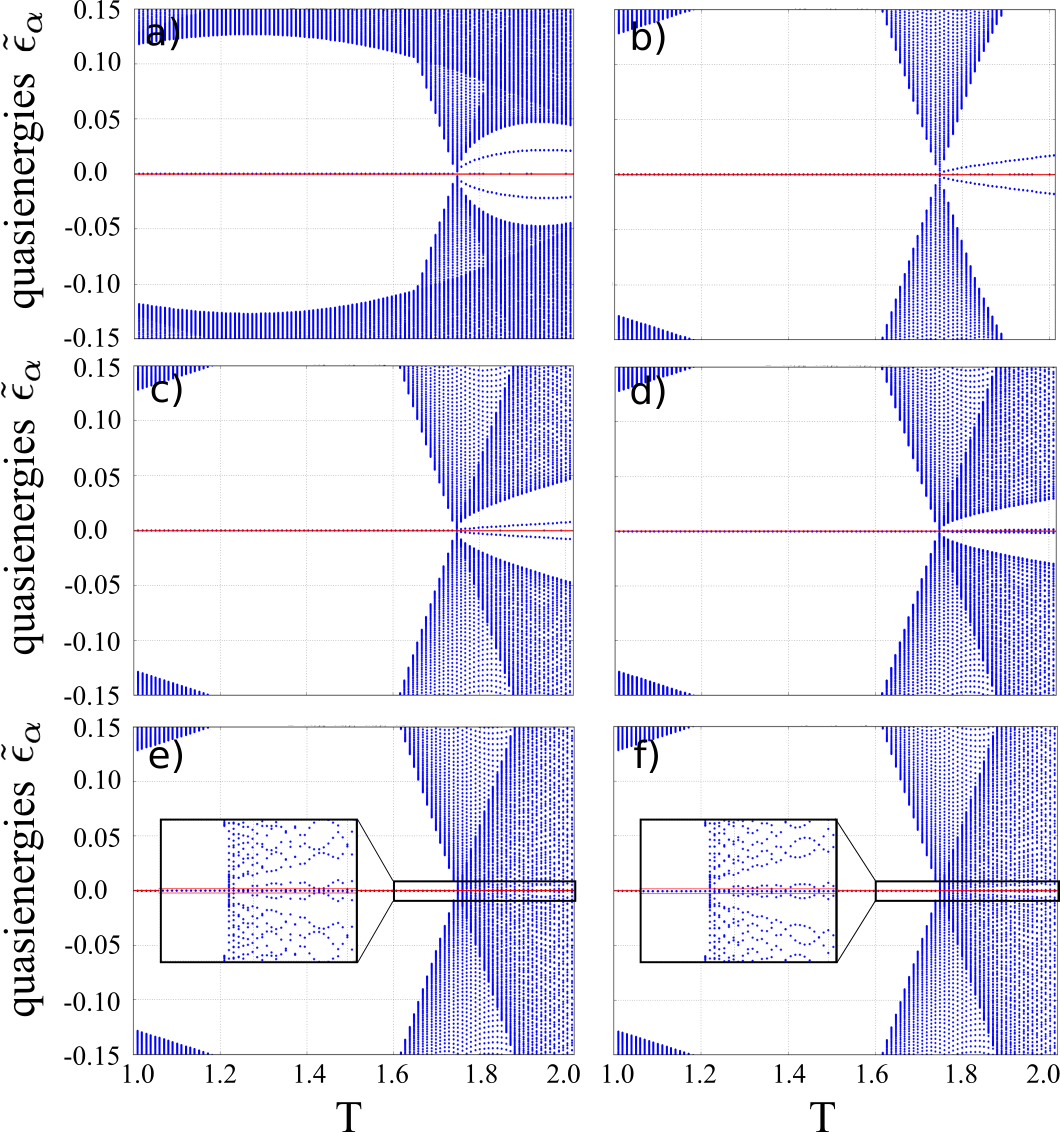}
\caption{Comparison of the quasienergy spectrum for the periodically driven Kitaev chain with single-driving protocol and with various commensurate double-driving protocols for a chain with $N=400$ fermions: a) single-driving protocol, b) 2:1 protocol, c) 3:1 protocol, d) 3:2 protocol, e) 4:3 protocol, f) 5:3 protocol. 
The red lines around 0 denote the threshold $10^{-3}$ used in the extraction of the phase diagram of figure \ref{fig:multifrequency-Kitaev-PD-comparison}.
The quasienergy spectrum was calculated as a function of the full period $T$ across a cut at $\mu_1 + \mu_2=0.7$.
The values of the static parameters are all $\tau=1.0$, $\Delta=0.1$, $\mu_0=0.1$.}
\label{fig:multifrequency-quasienergies-comparison-Kitaev}
\end{figure}

Because the form of the Floquet operator is strongly dependent on the commensuration scheme, the quasienergy spectrum will depend strongly on it, too. 
This finding is illustrated in Fig.~\ref{fig:multifrequency-quasienergies-comparison-Kitaev}, where the quasienergy spectrum around $\epsilon=0$ is compared across various commensuration schemes. 
Different driving protocols drastically alter the behavior of the quasienergy spectrum, specially compared to the single-frequency drive (panel a)).
In particular, the size and shape of the quasienergy gap is profoundly impacted by the type of driving protocol chosen, and so is the number of mid-gap states and quasienergies that correspond to MZMs.
A similar situation occurs also for gap closures at $\epsilon=\pi$ that are associated with appearance and disappearance of MZMs.

\begin{figure}
\centering
\includegraphics[width=\columnwidth]{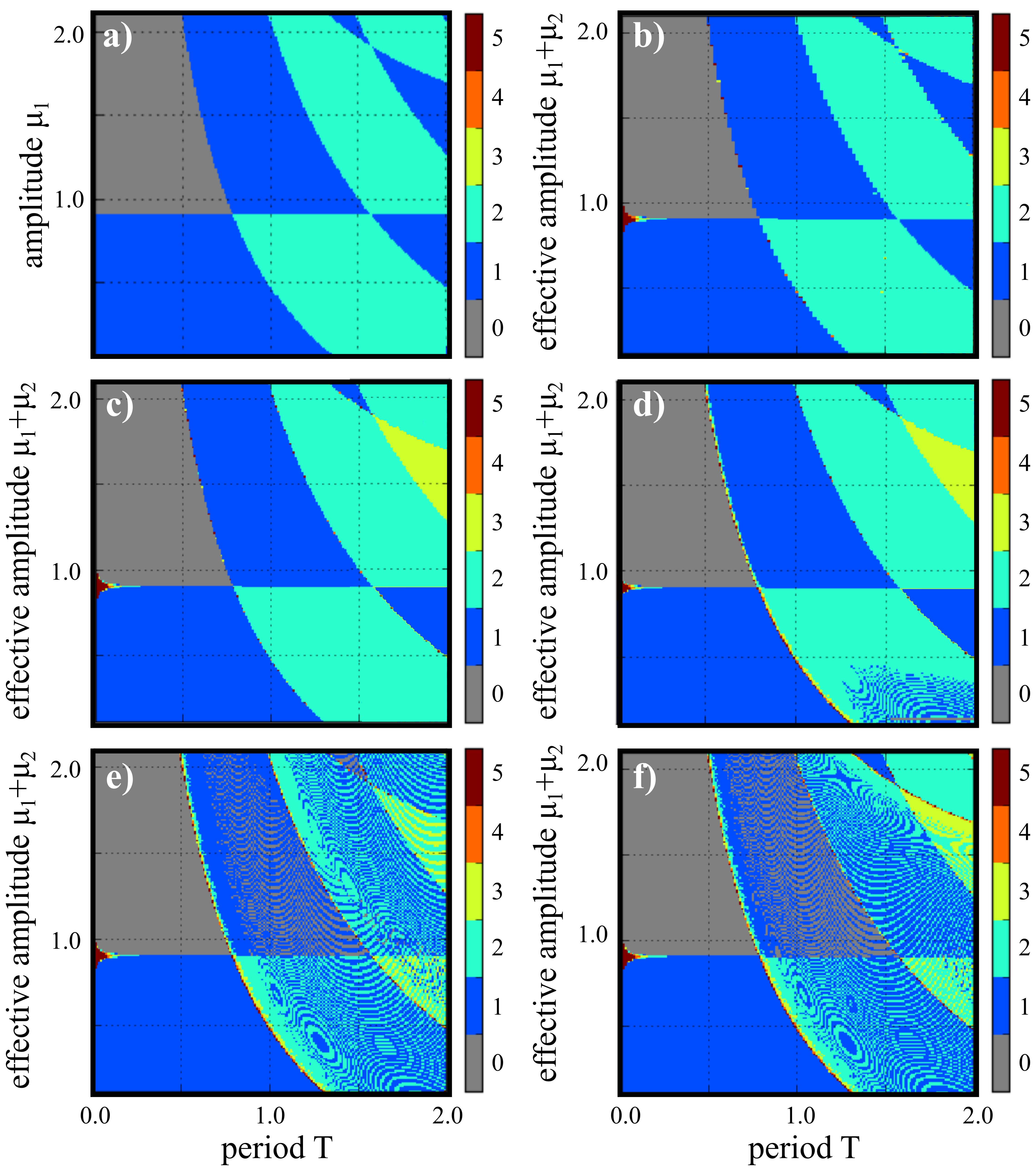}
\caption{Comparison of the phase diagram for the periodically driven Kitaev chain with single-driving protocol and with various commensurate double-driving protocols: a) single-driving protocol, b) 2:1 protocol, c) 3:1 protocol, d) 3:2 protocol, e) 4:3 protocol, f) 5:3 protocol.
At every point, the number of (asymptotic) FMMs was calculated from the quasienergy spectrum of a chain with $N=400$ fermions as the number of quasienergies $\epsilon$ within a threshold $10^{-3}$ from $0$ and $\pi$.
The values of the static parameters are all $\tau=1.0$, $\Delta=0.1$, $\mu_0=0.1$.
}
\label{fig:multifrequency-Kitaev-PD-comparison}
\end{figure}

Even though the quasienergy gap exhibits a great variation in its shape and size, from Fig.~\ref{fig:multifrequency-quasienergies-comparison-Kitaev} we can also appreciate how the various protocols all induce the same gap closure (at $T \approx 1.75$ for the chosen delta-drive protocol).
At the same time, the size of the quasienergy gaps and the number of Floquet-Majorana modes towards $\epsilon=0,\pi$ is different.
This finding is investigated more in depth in Fig.~\ref{fig:multifrequency-Kitaev-PD-comparison}, where the topological phase diagrams extracted from counting the number of $0$ and $\pi$ quasienergies are shown for the different driving protocols.
We can clearly see that, upon renormalizing the driving strength to the effective value $\mu_{\text{eff}} \equiv \mu_1 + \mu_2$, the TPTs all coincide with the ones observed for the single-driving case.
We note that the shaded patterns appearing in certain plots (d)-f)) are due to the quasienergy spectrum becoming more dense and the quasienergy gap becoming quite small (see also \ref{fig:multifrequency-quasienergies-comparison-Kitaev}).
This makes the arbitrary definition of the quasienergy threshold for the detection of zero and $\pi$ modes less precise.
Nevertheless, the patterns can be progressively reduced as $N \to \infty$, but this becomes computationally cumbersome.
Another numerical artefact is the appearance of the pockets around $T \approx 0$, $\mu_1+\mu_2\approx 0.9$ which seem to contain a large number of zero modes. 
Here the quasienergy spectrum is squeezed to very small values around zero because of the very small period $T$, leading to many quasienergies below the threshold for the definition of zero modes.
Both numerical artefacts  are observed also for other models (see next sections).

\begin{figure}
\centering
\includegraphics[width=\columnwidth]{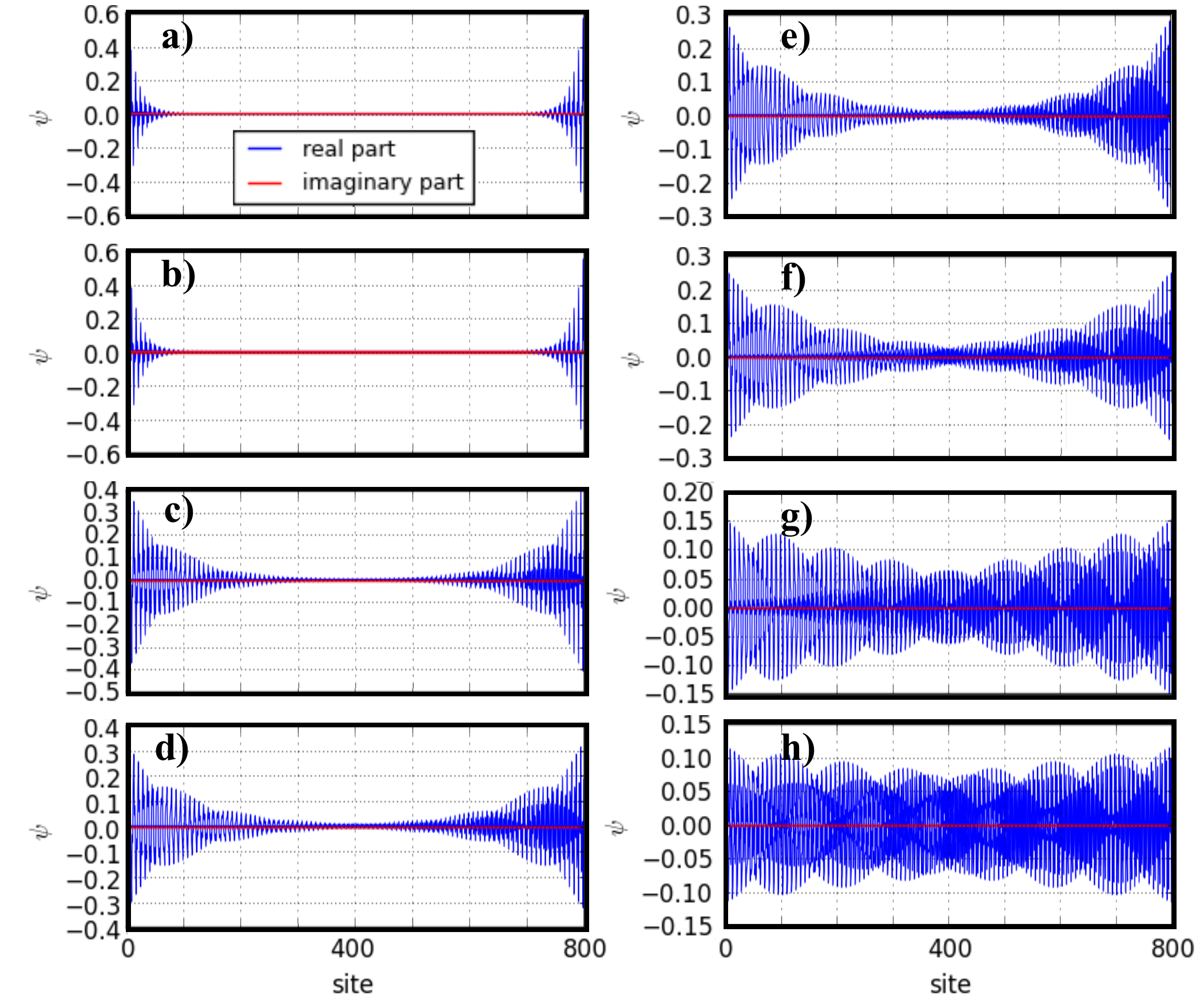}
\caption{Spatial extent of the same zero-quasienergy Floquet-Majorana mode at $T=2.0$, $\mu_1+\mu_2=1.5$, across eight different driving protocols. 
a) single-frequency driving ($\mu_1=1.5$),
b) 2:1 protocol,
c) 3:1 protocol,
d) 4:1 protocol,
e) 5:1 protocol,
f) 3:2 protocol,
g) 4:3 protocol,
h) 5:3 protocol.
For all multifrequency protocols $\mu_1=1.3$, $\mu_2=0.2$.
The values of the static parameters are all $\tau=1.0$, $\Delta=0.1$, $\mu_0=0.1$.}
\label{fig:edge-modes-Kitaev}
\end{figure}

To better illustrate how the different yet commensurate multifrequency driving affects the edge modes, in Fig.~\ref{fig:edge-modes-Kitaev} we plot a comparison across eight different driving protocols of the same Floquet MZM appearing at $T=2.0$, $\mu_1+\mu_2=1.5$ in the effective phase diagram.
From the plots we can appreciate how the localization of the edge modes (and therefore their overlap with fully delocalized bulk modes) is strongly influenced by the type of protocol used, up to the point where they can be rendered fully extended and completely lose their meaning of ``edge'' modes (Fig.~\ref{fig:edge-modes-Kitaev}g) and h) for 4:3 and 5:3 protocols).
This effect on the edge state localization could be useful in the dynamical manipulation of Floquet-Majorana modes, for instance in quantum computation schemes consisting of several segments of KCs longitudinally coupled with each other~\cite{Bauer:2018}: increasing the localization length for one segment could favor intra-segment coupling between the edge modes, while decreasing it could promote inter-segment interaction. 
The gap and the localization of the edge modes could be tuned at will by maintaining the same overall periodicity $T$, but splitting the drive into subpeaks that fulfill the sum rule \eqref{sum-rule-peaks} and the commensuration rule \eqref{commensuration-rule}.

\begin{figure}
\centering
\includegraphics[width=\columnwidth]{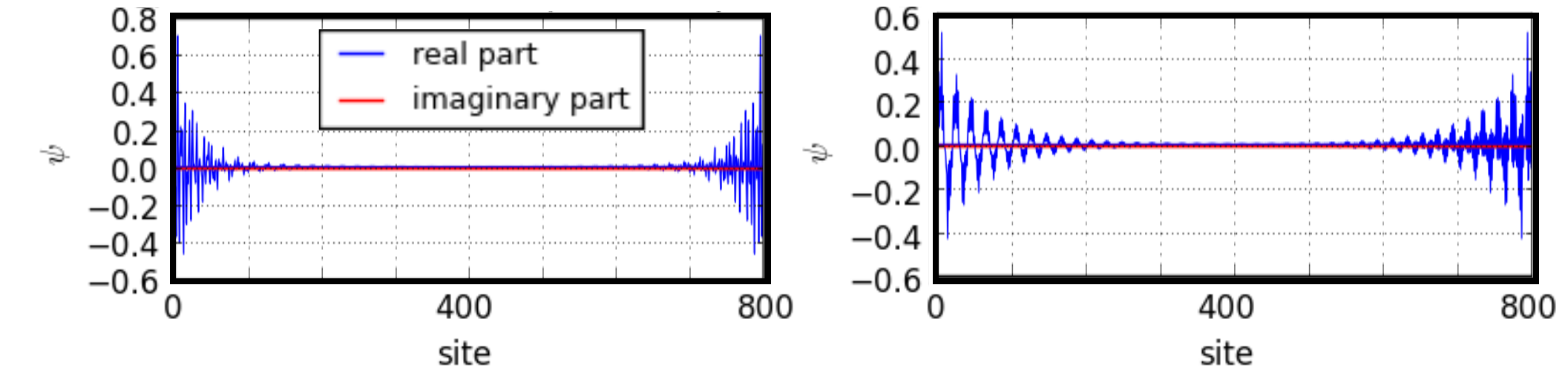}
\caption{Emerging new $\pi$ Floquet-Majorana edge states in the 3:1 driving protocol. The parameters are the same of Fig.~\ref{fig:edge-modes-Kitaev}.}
\label{fig:edge-modes-Kitaev-pi}
\end{figure}

The number of matrix exponential products in the Floquet operator for a $p_1:p_2$ driving protocol is $\mathcal{N}(p_1,p_2)=2(p_1+p_2-1)$ and can be used as estimation parameter of the complexity or ``richness" of the protocol itself.
For example, the richness of a $6:1$ protocol is comparable with the one of a $4:3$ protocol because $\mathcal{N}(6,1)=\mathcal{N}(4,3)=12$, while they are both more complex than a $2:1$ protocol with $\mathcal{N}(2,1)=6$.
Generally speaking, richer driving protocols appear to spread the quasienergies more uniformly across the spectrum and reduce the quasienergy gap to the edge states. 
This behavior in turn tends to delocalize the Floquet-Majorana modes away from the edges.
Exceptions to this trend however occur, as pictured in Fig.~\ref{fig:multifrequency-quasienergies-comparison-Kitaev}, where the gap around the MZM is larger in the 2:1 and 3:1 protocols than in the single-frequency case.
In fact, the number of edge modes can also be \emph{increased} by applying more complex multifrequency driving protocols.
This mechanism is illustrated in Fig.~\ref{fig:edge-modes-Kitaev-pi}, which depicts the generation of two additional edge-localized $\pi$-modes through the application of a 3:1 protocol.
Switching between different multifrequency protocols could therefore be a viable option to dynamically increase or reduce the number of edge modes without altering the overall periodicity of the drive.

\begin{figure}
\centering
\includegraphics[width=\columnwidth]{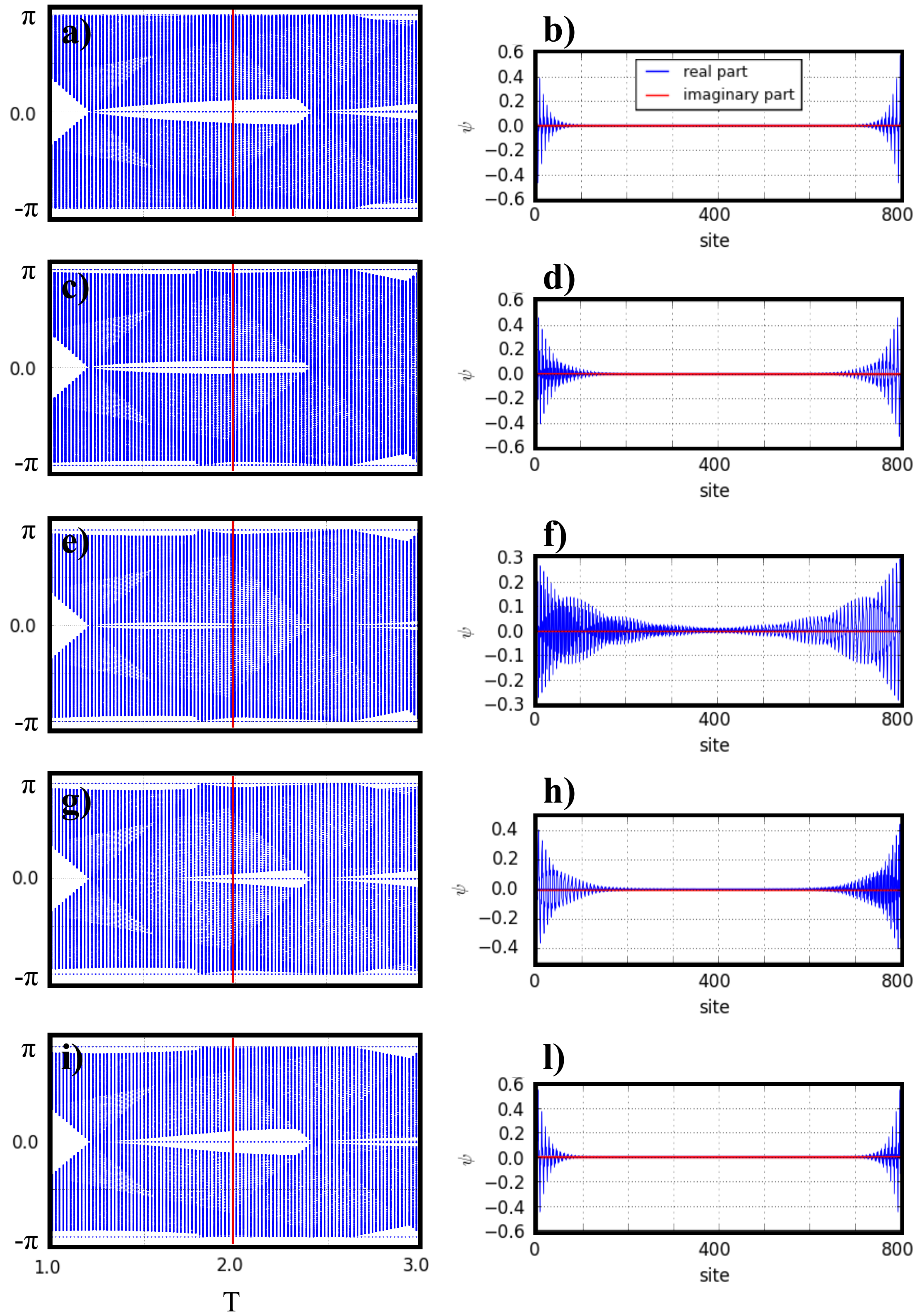}
\caption{Quasienergy spectrum (left panels) and zero-quasienergy Floquet-Majorana modes (right panels, corresponding to the vertical red line on the left) for the Kitaev chain of $N=400$ fermions, driven with the 2:1 multifrequency protocol at $T=2.0$ for different values of $\mu_1+\mu_2=1.5$.
a)-b) $\mu_1=1.4$, $\mu_2=0.1$.
c)-d) $\mu_1=1.0$, $\mu_2=0.5$.
e)-f) $\mu_1=0.8$, $\mu_2=0.7$.
g)-h) $\mu_1=0.5$, $\mu_2=1.0$.
i)-l) $\mu_1=0.1$, $\mu_2=1.4$.
The values of the static parameters are all $\tau=1.0$, $\Delta=0.1$, $\mu_0=0.1$.}
\label{fig:edge-modes-Kitaev-mu2}
\end{figure}

The localization of the edge modes can be tuned not only by operating with different driving schemes, but also within the same protocol using different distributions of the total driving strength $\mu_{\text{eff}}$ across the different drives.
This finding is shown in Fig.~\ref{fig:edge-modes-Kitaev-mu2}, where we compare the quasienergy spectrum and the corresponding localization of a zero-quasienergy Floquet-Majorana mode at $T=2.0$ (highlighted as a vertical red line). 
By varying the distribution of the total driving intensity across the two Dirac combs, the quasienergy spectrum and thus the size of the quasienergy gaps can be tuned at will (left panels).
Consequently, the localization length of the edge mode can be controlled (right panels), making it more or less susceptible to interactions with other bulk modes due to an increased wave function overlap.
Note that the transfer of driving intensity betw
een the two pulses could be achieved adiabatically and thus does not introduce quench dynamics.

Our results show that the additional driving amplitudes offer a new dynamical dimensionality to explore a wider range of localization lengths in parameter space, avoiding critical fluctuations of the static transition boundaries.
Experimentally, this is significant because dynamical changes to the stability of the edge modes can be obtained without changing the static parameters of the system, \textit{i.e} with the same static setup.
This could be useful when the tuneable range of the hardwired static parameters is small because of technical challenges.
Additionally, certain novel topological transport protocols~\cite{Bello:2016,Perez-Gonzalez:2019} inherently require to operate in driven settings.
In these cases, an additional lever to control the localization of the Floquet modes could be important, specially when more than one mode localizes at the ends and there is transport between more channels.

The fact that different multifrequency drives realize the same effective topological phase diagram can be understood by applying the general considerations of section~\ref{sec:multifreq}.
Let us generally consider a commensurate $p:q$ drive with delta kicks of strength $\mu_1 T_1$ and $\mu_2 T_2$ such that $p T_1 + q T_2 = T$, such as Eq.~\eqref{mu-double-frequency-Kitaev-delta}.
Because the topology will be governed by the closure or opening of the quasienergy gap at the HSPs, we can think of evaluating the Hamiltonian $\mathcal{H}_{KC,k}(t)$ around the HSPs $k_0=0,\pi$, which will eliminate to $\sigma^y$ term due to $\sin k_0 = 0$.
The Floquet operator then becomes a product of diagonal matrices that commute with each other, and can be written in a compact form. For example, for a 3:2 drive:
\begin{widetext}
\begin{align}
U_{k_0}(T,0) &= 
\exp \left( 2i (\mu_1 T_1 + \mu_2 T_2) \sigma^z \right) 
\exp \left(-2i (\tau \cos k_0 - \mu_0) T_1 \sigma^z \right) \exp \left( 2i (\mu_1 T_1) \sigma^z \right) \exp \left(-2i (\tau \cos k_0 - \mu_0) (T_2-T_1) \sigma^z \right) \nonumber \\
& \qquad \times \exp \left( 2i (\mu_2 T_2) \sigma^z \right) 
\exp \left(-2i (\tau \cos k_0 - \mu_0) (T_2-T_1) \sigma^z \right) \exp \left( 2i (\mu_1 T_1) \sigma^z \right) 
\exp \left(-2i (\tau \cos k_0 - \mu_0) T_1 \sigma^z \right) \nonumber \\
&= \exp \left[ -2iT [\tau \cos k_0 - (\mu_0 + \mu_1 + \mu_2)] \sigma^z \right] \nonumber \\
&\equiv \exp \left[ -2i T(\tau \cos k_0 - (\mu_0 + \mu_{\text{eff}}) )  \sigma^z \right], 
\label{floquet-operator-k0-general}
\end{align}
\end{widetext}
where we have used the commensuration condition $3 T_1 = 2 T_2 = T$ and defined $\mu_{\text{eff}} \equiv  \mu_1 + \mu_2$.
The Floquet operator for the single-frequency case is
\begin{align}
U_{k_0}(T,0) &= \exp \left[ 2i (\mu_1 T) \sigma^z \right] \exp \left[ -2i T (\tau \cos k_0 - \mu_0 ) \sigma^z \right] \nonumber \\
&= \exp \left[ -2iT( \tau \cos k_0 - (\mu_0 + \mu_1)) \sigma^z \right]
\end{align}
and has precisely the same form in parameter space as the multifrequency driving case, provided $\mu_1$ is replaced with $\mu_{\text{eff}}$. 
Therefore, the effective topological phase boundaries of the two driving schemes must coincide.
The above calculation can be generalized to the commensuration condition $p T_1 = q T_2 = T$ to obtain the same result.
More generally, this result is valid also for multifrequency delta-drives with an arbitrary number of commensurate periods $T_1, T_2, T_3, \dots$ satisfying $p_1 T_1 = p_2 T_2 = p_3 T_3 = \cdots = T$, where $T$ is the smallest common multiple of every sub-period $T_i$, and $p_i$ is the corresponding multiplicity
Because the gap can only close at the HSPs $k_0=0,\pi$, all possible combinations of multifrequency delta-drive are collapsed to the same topological phase diagram, provided that
\begin{equation}
 \mu_1 \to \mu_{\text{eff}} = \sum_{i=1} \mu_i
\label{sum-rule-peaks}
\end{equation}
While the gap closures coincide at the HSPs, slightly away from them the effect of the $\sigma^y$ term will modify the quasienergy spectrum and strongly influence the shape and size of the gap as we tune away from criticality.

The precise form of the topological phase boundaries can also be read out of the quasienergy dispersion at the HSPs,
\begin{align}
\theta(k_0) &\equiv \arccos \left[ \Tr \left[ U_{k_0}(T,0) \right] /2 \right] \nonumber \\
&= \arccos \left[ \cos \left( -2( \tau \cos k_0 - (\mu_0 + \mu_{\text{eff}})T ) ) \right) \right] \nonumber \\
&=  -2T( \tau \cos k_0 - (\mu_0 + \mu_{\text{eff}}) ).
\end{align}
By demanding that the gap closes, $\theta(k_0)=m \pi$, $m \in \mathbb{Z}$ (the multiples of $\pi$ stem from folding back all the Floquet bands to 0 or $\pi$), we obtain
\begin{equation}
\mu_0 + \mu_{\text{eff}} = \pm \tau  - \frac{m\pi}{2T}, 
\label{condition-multifrequency-delta}
\end{equation}
which precisely coincides with the topological phase boundaries.


\section{SSH model}
\label{sec:SSH}

We now consider the SSH chain belonging to the BDI class.
The SSH chain describes non-interacting, spinless fermions hopping on a 1D lattice of $N$ unit cells with staggered hopping amplitudes between two sublattices $A$ and $B$. 
Each unit cell $n$ consists of two sites, one on sublattice A described by operator $a_n$, and one on sublattice B described by operator $b_n$.
The dynamics of the system is described by a Hamiltonian of the form
\begin{align}
\mathcal{H}_{SSH} &= \sum_{n=1}^{N} \left[  \left( v a_n^{\dagger} b_n + u_0 a_{n+1}^{\dagger} b_{n} + h.c.  \right) \right. \nonumber \\
& \qquad \qquad + \left.  M \left( a_n^{\dagger} a_n - b_n^{\dagger} b_n \right) \right].
\end{align}
Because of this staggered configuration of different intra- and inter-cell hopping, the SSH with $M=0$ is sometimes used to model the structure of polyacetilene molecules~\cite{su79}, while for $M \neq 0$ it can model diatomic polymers~\cite{Rice:1982}.
In the following, we will set $M=0$ for simplicity.
For $M = 0$ and $u_0>v$ the system has a nontrivial topology with two (fermionic) edge states appearing at the end of the chain, one on sublattice $A$ and the other on sublattice $B$.

Once again, because of the bulk-edge correspondence, we can examine the topology also in a ring geometry.
The BdG bulk Hamiltonian of the SSH model can be obtained by Fourier transforming to momentum space in a procedure similar to the one highlighted in the previous section, yielding~\cite{Asboth-book}
\begin{equation}
\mathcal{H}_{SSH,k} = 
(v + u_0 \cos k ) \sigma^x + u_0 \sin k \sigma^y
\end{equation}
with an energy dispersion
\begin{equation}
E_{SSH}(k) = \pm \sqrt{ v^2 + u_0^2 + 2vu_0 \cos k},
\label{ssh-static-dispersion}
\end{equation}
which exhibits a gap closure at $k_0 = \pm \pi$ for $v = u_0$, concomitantly with the expected TPT in the open chain.


Analogously to what already discussed for the Kitaev chain, periodic driving applied on the SSH chain can also induce edge-localized modes associated with quasienergy $\pm \pi$~\cite{Asboth:2014,Fruchart:2016,Bello:2016,Niklas:2016,Balabanov:2017,Perez-Gonzalez:2019}.
These $\pi$ modes have already been experimentally detected, for instance in coupled corrugated waveguides~\cite{Cheng:2019}.
As opposed to the static case, in the driven case the gap closures in the dispersion can also occur at the other HSP, $k_0=0$.

\subsection{Multifrequency driving in the SSH model}

Multifrequency driving can be applied here too to control the size of the quasienergy gaps and the number of topological excitations in each phase, all without changing the location of the gap closures in the phase diagram that delineate the TPTs. 
We vary the intercell hopping as
\begin{align}
u_0 \to u(t) &= u_0 + u_1 T_1 \sum_{m \in \mathbb{Z}} \delta(t - mT_1) \nonumber \\
& \qquad + u_2 T_2 \sum_{m' \in \mathbb{Z}} \delta(t - m' T_2),
\end{align}
where the commensuration condition $p_1 T_1 = p_2 T_2 = T$ holds.
We then probe the topology as a function of the driving parameters $u_1$, $u_2$, and $T$ by counting the quasienergies at $\epsilon T=0, \pm \pi$ extracted from the stroboscopic Floquet Hamiltonian.
Note that for the single-driving case we obtain
\begin{align}
&U_{SSH,k}(T,0) = \exp \left[ -i T \left( u_1 \cos k \sigma^x + u_1 \sin k \sigma^y \right) \right] \nonumber \\
& \quad \times   \exp \left[ -i T \left( (v + u_0 \cos k) \sigma^x + u_0 \sin k \sigma^y \right) \right].
\label{Floquet-op-ssh-single-driving}
\end{align}

\begin{figure}
\centering
\includegraphics[width=\columnwidth]{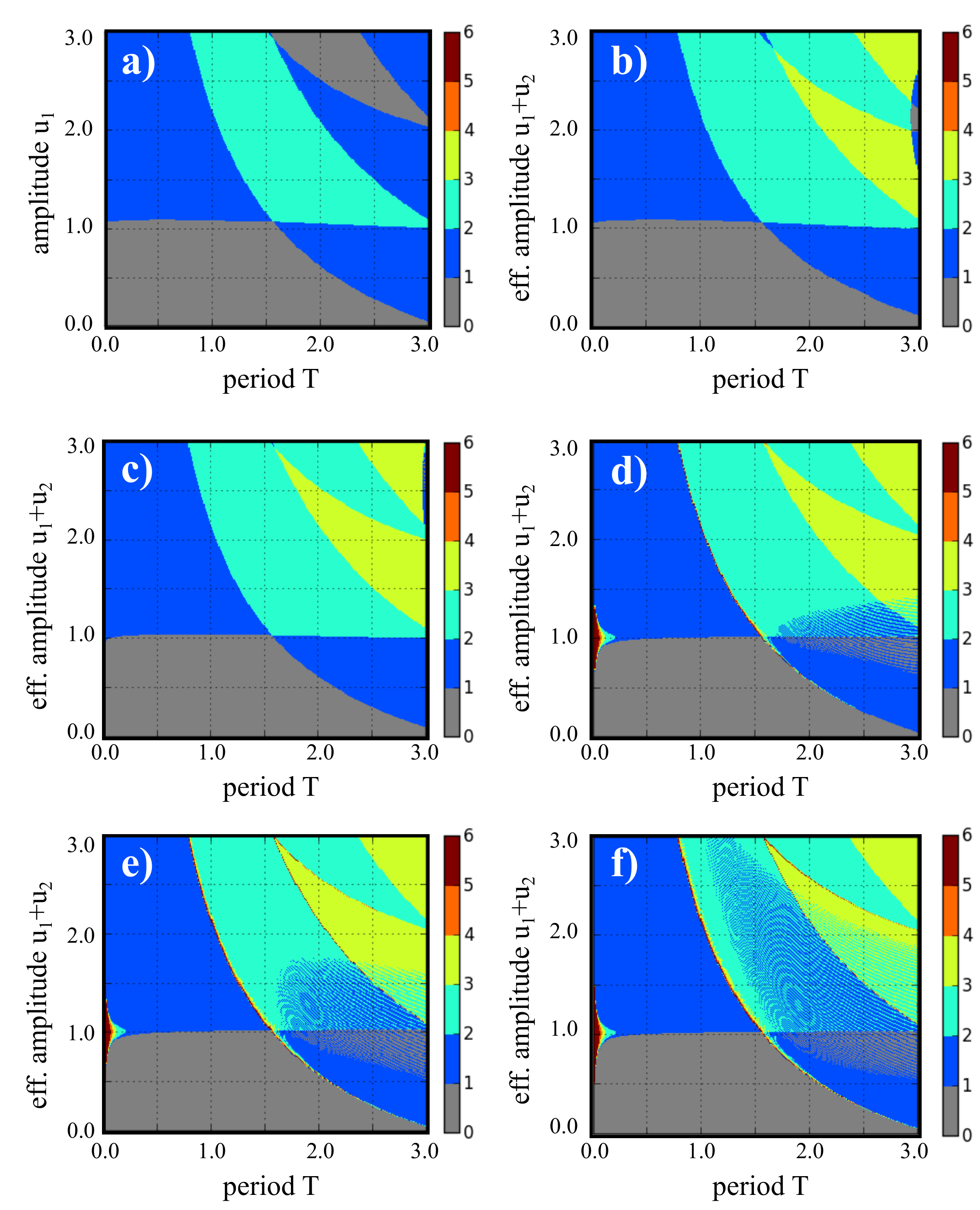}
\caption{Comparison of the topological phase diagrams of the periodically driven SSH chain with the single-driving protocol and with various commensurate double-driving protocols for a chain of $N=100$ unit cells: a) single-driving protocol, b) 2:1 protocol, c) 3:1 protocol, d) 3:2 protocol, e) 4:3 protocol, f) 5:3 protocol.
The phase diagram was obtained by counting the number of 
quasienergies within a certain threshold $\delta$ from $0$ and $\pi$. For a)-c) $\delta=10^{-3}$, while for d)-f) $\delta=10^{-2}$.
The values of the static parameters are all $v=1.0$, $u_0=0.0$.
}
\label{fig:comparison-PD-SSH}
\end{figure}

Fig.~\ref{fig:comparison-PD-SSH} presents a comparison of the topological phase diagrams obtained for the single-frequency and various double-frequency protocols. 
From the figure we can clearly see that all standard TPTs occur at the same position in phase space for every driving scheme, provided that we account for the renormalization $u_1 \to u_{\text{eff}} = u_1 + u_2$ in the driving strength.
An exception is provided by TPTs at non HSPs associated with quasienergy backfolding, similar to the case ``frozen dynamics" seen in the periodically driven Kitaev chain.
This phenomenon is however dependent on the particular choice of delta-drive and is not universal.
This is briefly discussed in appendix \ref{app:frozen-SSH}.

We can understand the backfolding of the phase diagrams from the double-frequency to the single-frequency case once again by considering the form of the Floquet operator at the gap closures at $k_0=0,\pm\pi$:
\begin{align}
U_{k_0}(T,0) 
&= \exp \left[ -iT  \left( v + (u_0 + u_1 + u_2) \cos k_0 \right) \sigma^x \right] \nonumber \\
&= \exp \left[ -iT  \left( v + (u_0 + u_{\text{eff}}) \cos k_0 \right) \sigma^x \right],
\end{align}
which has the same form of the Floquet operator at $k=k_0$ for the single driving, Eq. \eqref{Floquet-op-ssh-single-driving}, provided $ u_1 \to u_{\text{eff}}$.

\begin{figure}
\centering
\includegraphics[width=\columnwidth]{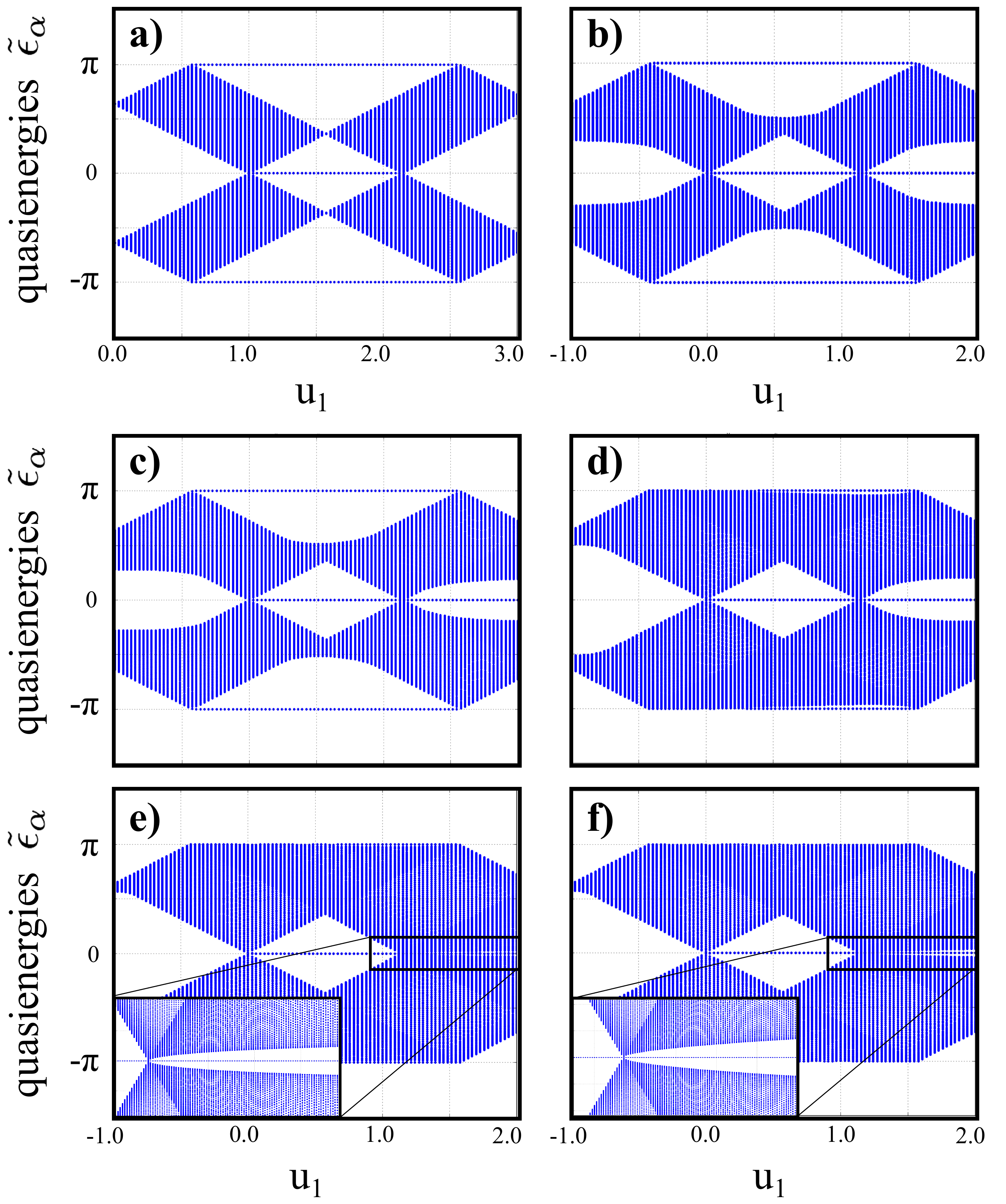}
\caption{Comparison of the quasienergy spectra of the periodically driven SSH chain with the single-driving protocol and with various commensurate double-driving protocols for a chain of $N=100$ unit cells and $T=2.0$: a) single-driving protocol, b) 2:1 protocol, c) 3:1 protocol, d) 3:2 protocol, e) 4:3 protocol, f) 5:3 protocol.
The values of the static parameters are all $v=1.0$, $u_0=0.0$, while the driving strength of the second pulse was chosen to be $u_2=1.0$ (b-f), such that $u_{\text{eff}} = u_1 + u_2 \in [0,3]$ for comparison with the single-driving protocol.}
\label{fig:multifrequency-quasienergies-comparison-SSH}
\end{figure}

Even though the TPTs remain consistent throughout the effective phase space, the number of edge modes and their stability depend heavily on the driving protocol used, and so does the size of the quasienergy gaps at $0$ and $\pi$.
These characteristics can be ascertained from Fig.~\ref{fig:multifrequency-quasienergies-comparison-SSH}, where the quasienergy spectrum across $T=2.0$ is plotted for various driving protocols.
In particular, it is interesting to also note that multifrequency driving stabilizes additional zero-energy modes for $u_1 \gtrsim 1.2$ that are not present in the single-frequency case.
As in the KC, multifrequency driving in the SSH model is used to modify the number and localization of the edge modes while maintaining the overall shape of the phase diagram  (\textit{i.e.} the location of the gap closures).



\begin{figure}
\centering
\includegraphics[width=\columnwidth]{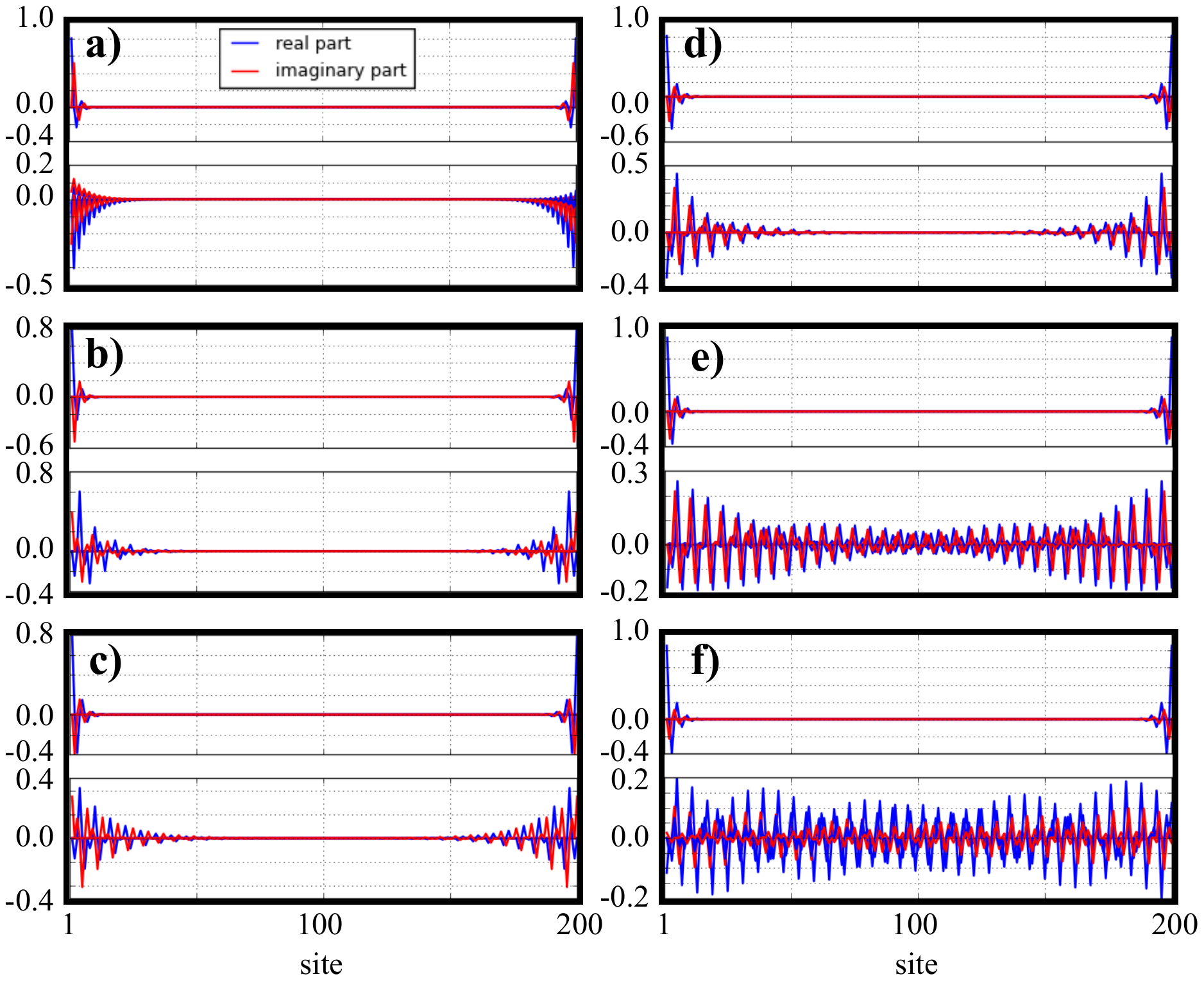}
\caption{Edge mode localization in the driven SSH model with $N=100$ unit cells, across six different driving protocols with $T=2.0$.
For each panel, the upper plot depicts a zero-quasienergy mode, while the lower plot corresponds to a $\pi$-quasienergy mode.
a) single-frequency driving ($u_1=2.0$),
b) 2:1 protocol,
c) 3:1 protocol,
d) 4:1 protocol,
e) 3:2 protocol,
f) 4:3 protocol.
For all multifrequency protocols $u_1=1.8$, $u_2=0.2$.
The values of the static parameters are all $v=1.0$, $u_0=0.0$.}
\label{fig:edge-modes-ssh}
\end{figure}

Like in the KC, control over the edge modes is achieved also by tuning the size of the quasienergy gaps dynamically over different driving protocols. 
As a consequence, a different localization of the edge modes and a different degree of hybridization among themselves can be obtained.
This result is depicted in Fig.~\ref{fig:edge-modes-ssh}, where the edge mode localization at $T=2.0$ and $u_{\text{eff}}$ is plotted for various driving protocols.
In the figure we can clearly see how, for the parameters chosen, the localization length of the $\pi$-mode can be manipulated over a very large interval: from strong localization at the edges in the single-frequency protocol, to complete hybridization in the 4:3 protocol.
It is interesting to note that, while the $\pi$ mode localization is controlled, the zero edge mode remains strongly localized across all the different protocols.
This finding therefore illustrates also that zero and $\pi$-mode localization can be tuned independently from each other by choosing appropriate regimes and driving protocols.

\section{Creutz ladder}
\label{sec:Creutz}

The third topological model analyzed in this paper is a Creutz-type ladder consisting of two 1D sublattices (A and B) of $N$ sites each~\cite{Creutz:1999}.
This model belongs to yet another symmetry class --- the complex symmetry class AIII --- and possesses a topological phase diagram that is equivalent to the one of the Haldane model, and can be therefore be viewed as its 1D counterpart~\cite{Gholizadeh:2018}.
Spinless electrons hop along the two sublattice with opposite strengths $t_x$ and $-t_x$.
The lattices are additionally coupled to each other via nearest-neighbor (NN) hoppings $t_y$ and next-nearest-neighbor (NNN) hoppings $t_{xy}$.
Finally, a magnetic field perpendicular to the ladder plane induces an additional Peierls phase $\theta$ along the NNN bonds.
The full Hamiltonian of the system is therefore
\begin{align}
\mathcal{H}_{CL} &=\sum_{n=1}^N \left[ t_x \left( b_n^{\dagger} b_{n+1} - a_n^{\dagger} a_{n+1} \right) - t_y a^{\dagger}_n b_n \right. \nonumber \\
&  \qquad  \left.- t_{xy} \left( b_n^{\dagger} a_{n+1} e^{i \theta} + a_n^{\dagger} b_{n+1} e^{-i \theta} \right) + h.c. \right],
\label{Ham-CL}
\end{align}
where $a^{(\dagger)}_n$ and $b^{(\dagger)}_n$ denote creation/annihilation operators at site $n$ for the sublattice A and B, respectively.
In the following, we will set $t_x=t_y=t$ for simplicity, and normalize the Hamiltonian in units of $t_{xy}$, defining an effective hopping parameter $m \equiv \frac{t}{t_{xy}}$.
This type of Creutz ladder can be realized in ultracold fermionic systems in optical lattices, where the NNN hopping can be engineered via two-photon resonant coupling between the orbitals by periodic shaking the lattice~\cite{Bermudez:2009,Mazza:2012,Juenemann:2017}, or in photonic systems~\cite{Zurita:2019}.


In the static Creutz ladder with open boundary conditions, topological interference effects lead to the appearance of zero-energy edge states within the regions delimited by the curves $m = \pm 2 \sin \theta$.
With periodic boundary conditions, it is possible to describe the topology from the bulk Hamiltonian
\begin{align}
\frac{\mathcal{H}_{CL,k} }{t_{xy}} &= \left[ (m + 2 \cos(k+ \theta)) \: \sigma^x + 2 m \cos(k) \: \sigma^z \right].
\label{CL-bulk-Ham}
\end{align}
%
Along the curves $m = \pm 2 \sin \theta$ that delineate the static topological phase transitions, the bulk energy dispersion exhibits Dirac-type gap closures at $k=\pm \frac{\pi}{2}$.

%

\subsection{Multifrequency driving in the Creutz ladder}

As in the previously discussed models, it is possible to drive the parameters of the Creutz ladder with multiple commensurate frequencies to obtain out-of-equilibrium topological phases.
In the following, we concentrate on the multifrequency drive applied to the parameter $m$, again in the form of delta kicks
\begin{align}
m \to m(t) &= m_0 + m_1 T_1 \sum_{n \in \mathbb{Z}} \delta(t - n T_1)  \nonumber \\
& \qquad + m_2 T_2 \sum_{n' \in \mathbb{Z}} \delta(t - n' T_2),
\end{align}
with the usual commensuration condition $p_1 T_1 = p_2 T_2 = T$. 
Once again, we probe the stroboscopic topology by constructing the Floquet effective Hamiltonian \eqref{eq:Floquet-operator} and analyzing the appearance and disappearance of zero and $\pi$ modes as a function of the driving parameters $m_i$ and the period $T$.
As a reminder, for the single-driving case, the Floquet operator takes the following form
\begin{align}
U_k(T,0) &= \exp \left[ -i( 2 m_1 T \cos k \: \sigma^z + m_1 T \: \sigma^x) \right] \nonumber \\
& \quad \times \exp \left[ -i T ( 2 m_0 \cos k \: \sigma^z +  \right. \nonumber \\
& \qquad \qquad \quad  \: \left.  (m_0 + 2 \cos(k + \theta) ) \: \sigma^x) \right]
\end{align}
while for multifrequency driving the Floquet operator involves the product of more exponentials.

\begin{figure}[h!]
\centering
\includegraphics[width=\columnwidth]{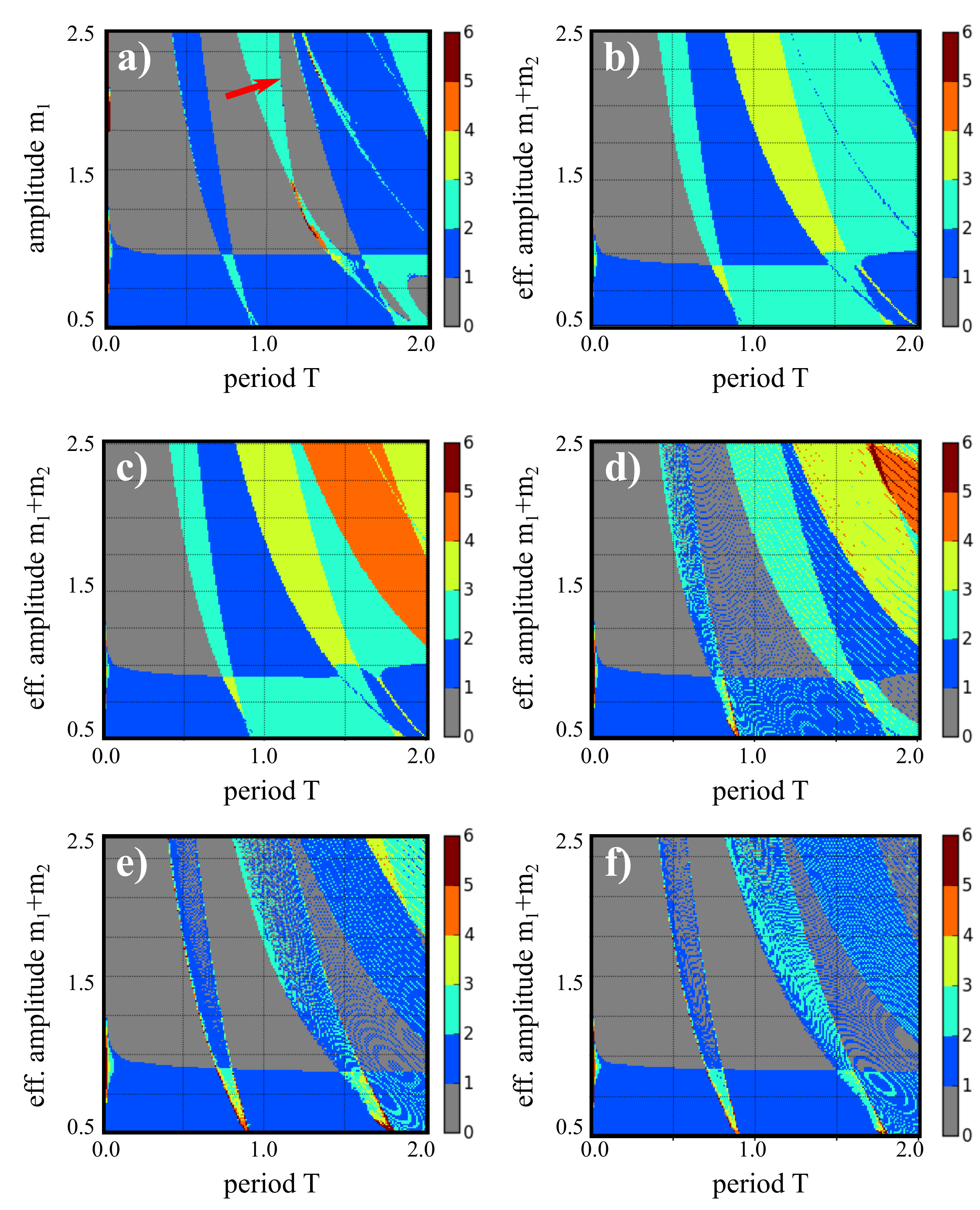}
\caption{Comparison of the topological phase diagrams of the periodically driven Creutz ladder with the single-driving protocol and with various commensurate double-driving protocols for a chain of $N=200$ ladder rungs: a) single-driving protocol, b) 2:1 protocol, c) 3:1 protocol, d) 3:2 protocol, e) 4:3 protocol, f) 5:3 protocol.
The phase diagram was obtained by counting the number of (asymptotic) quasienergies within a certain threshold $\delta$ from $0$ and $\pi$. For a)-c) $\delta=10^{-3}$, while for d)-f) $\delta=10^{-2}$.
The values of the static parameters are all $m_0=1.0$, $\theta=\pi/2$.
}
\label{fig:PD-creutz}
\end{figure}

\begin{figure}
\centering
\includegraphics[width=\columnwidth]{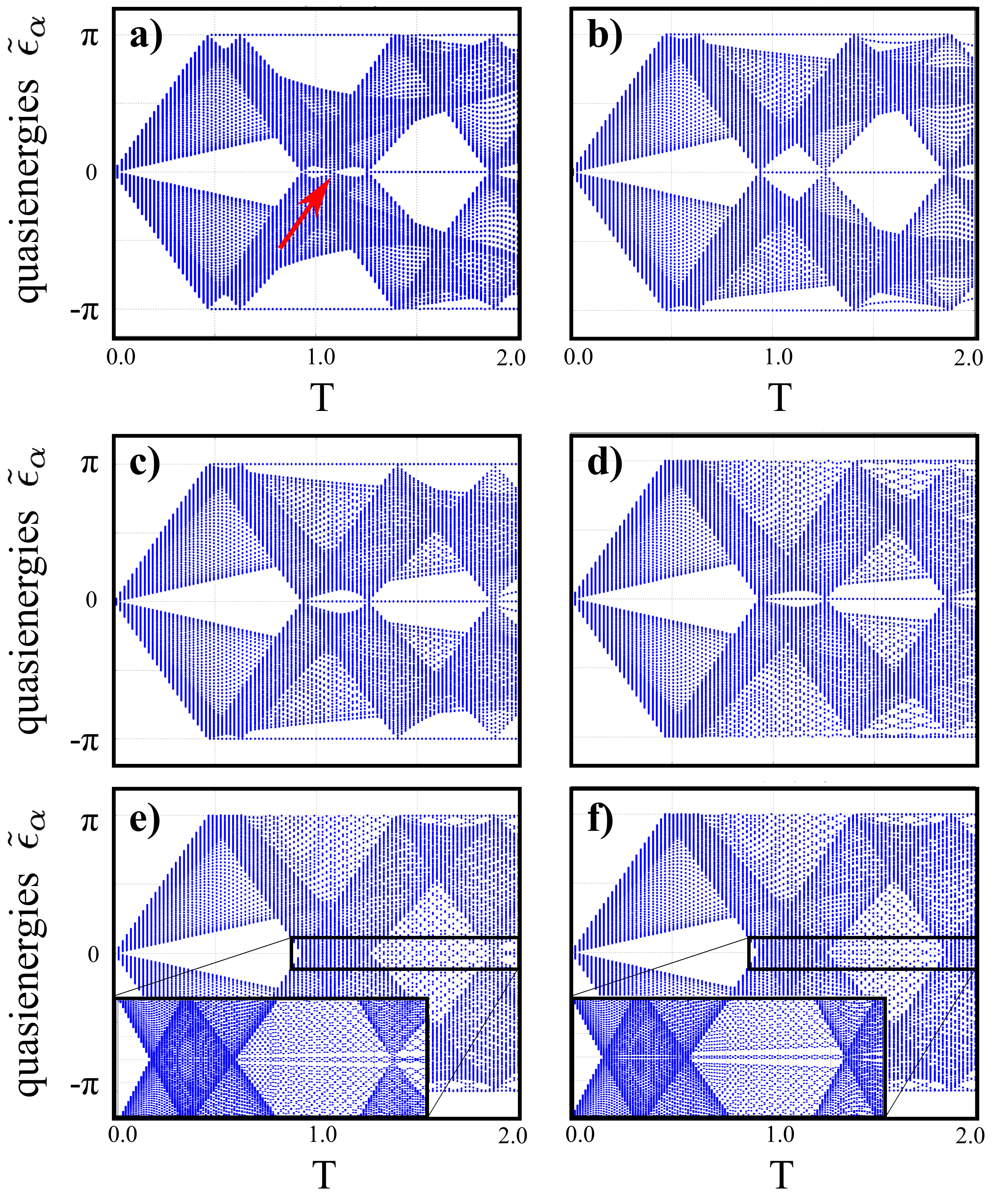}
\caption{Comparison of the quasienergy spectra of the periodically driven Creutz ladder with the single-driving protocol and with various commensurate double-driving protocols for a chain of $N=200$ fermions: a) single-driving protocol, b) 2:1 protocol, c) 3:1 protocol, d) 3:2 protocol, e) 4:3 protocol, f) 5:3 protocol.
The values of the static parameters are all $m_0=0.0$, $\theta = \frac{\pi}{2}$, while the intensities of the drives were chosen to be $m_1+m_2=2.0$.}
\label{fig:multifrequency-quasienergies-comparison-creutz}
\end{figure}

The stroboscopic topological phase diagrams obtained for the Creutz ladder are displayed in Fig.~\ref{fig:PD-creutz} for the single-frequency protocol (panel a)) and various multifrequency protocols (panels b)-f)).
Like in the previous systems, for all protocols considered, a visual comparison reveals that the topological phase boundaries have the same shape in the effective parameter space, apart from minor distortions due to finite size effects and the arbitrary choice of the quasienergy threshold.
The only notable exception is the presence of a seemingly additional TPT in the single-driving protocol (marked with a red arrow in Figs.~\ref{fig:PD-creutz}a) and \ref{fig:multifrequency-quasienergies-comparison-creutz}a)), which is absent for the other driving protocols.
However, as explained in Appendix \ref{app:frozen-CL}, this transition stems from frozen dynamics: it corresponds to an additional gap closure at a non-HSP $k_0 \neq 0, \pm \pi/2, \pm \pi$ due to the particular choice of delta-driving.
Since transitions associated with frozen dynamics are in general neither systematic nor robust, we will ignore them in our general discussion.

Fig.~\ref{fig:multifrequency-quasienergies-comparison-creutz} depicts the shape of the quasienergy spectrum for the different driving protocols along the same cut with $m_{\text{eff}}=2.0$.
As seen in the previous systems belonging to different symmetry classes, despite the gap closures and the TPTs occur at the same points, the quasienergy spectrum greatly differs from case to case.
This affects both the number of $0$ and $\pi$ edge modes present in each phase, and the size of the gap separating them from bulk modes.
By dynamically switching between different driving protocols, it should therefore be possible to control the stability of such edge modes like in the other models.

The equivalence of the topological phase diagrams for different multifrequency driving protocols can, once again, be traced back to the fact that the TPTs are governed by the gap closures at a handful of HSPs.
For the Creutz ladder, the HSPs are located at $k_0 = \pm \frac{\pi}{2}$.
At these points, because $\cos k_0 = 0$, the instantaneous Hamiltonian is given only by the $\sigma^x$ component and the time-ordering in the Floquet operator can be dropped, yielding
%
%
\begin{align}
U_{k_0}(T,0) 
 &= \exp \bigg[ -i \bigg( (m_0 + 2 \sin k_0 \sin \theta) T \nonumber \\
& \qquad \qquad +  (m_1 + m_2) T \bigg) \sigma^x \bigg].
\end{align}
Precisely like in the previous cases, we can then define an effective parameter $m_{\text{eff}} \equiv m_1 + m_2$.
Because the Floquet operator for the multifrequency protocols written in terms of the effective parameter has the same form as the one for the single-frequency protocol, the TPTs in both cases must coincide.

\begin{figure}
\centering
\includegraphics[width=\columnwidth]{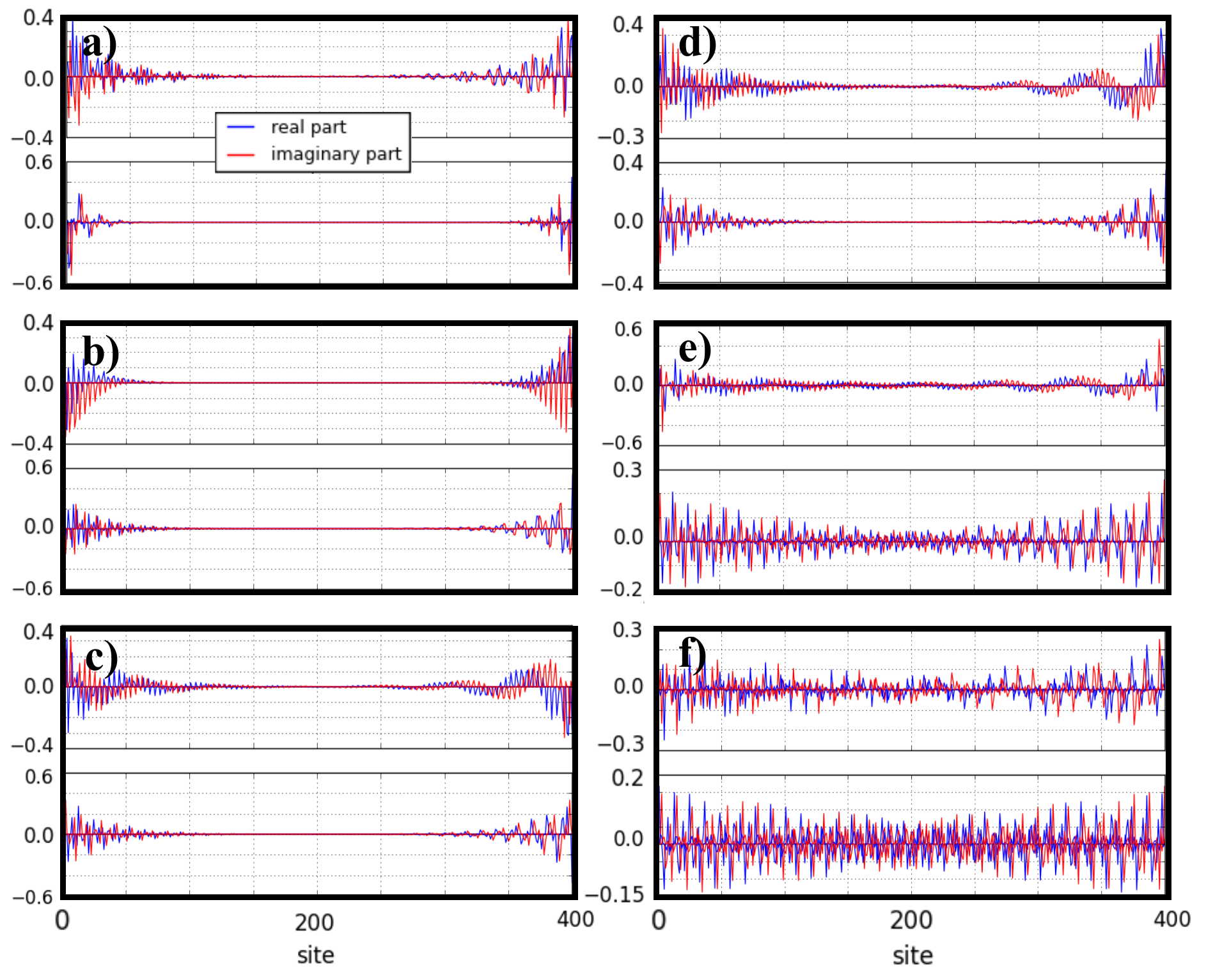}
\caption{Edge mode localization in the driven Creutz ladder with $N=200$ rungs, across six different driving protocols with $T=1.5$.
For each panel, the lower plot depicts a zero-quasienergy mode, while the upper plot corresponds to a $\pi$-quasienergy mode.
a) single-frequency driving ($m_1=2.0$),
b) 2:1 protocol,
c) 3:1 protocol,
d) 4:1 protocol,
e) 3:2 protocol,
f) 4:3 protocol.
For all multifrequency protocols $m_1=1.8$, $m_2=0.2$.
The values of the static parameters are all $m_0=0.0$, $\theta=\pi/2$.}
\label{fig:edge-modes-creutz}
\end{figure}

Like in the other models, the different size of the quasienergy gaps obtained with different driving protocols is reflected in a different edge mode localization.
This result is illustrated in Fig. \ref{fig:edge-modes-creutz}, where the edge modes at $T=1.5$, $m_{\text{eff}}=2.0$ are calculated for various driving protocols.
From the figure, we can appreciate that the edge modes can be made to localize or delocalize over a great range of distances, often with the presence of an oscillatory tail typical of Floquet systems~\cite{Thakurathi:2013}.
Overall, more complex driving protocols tend to fill up the quasienergy spectrum more evenly, and lead to a stronger delocalization of the edge modes.
This change can be quite dramatic, to the point where a full hybridization between the edge modes occurs (panels e) and f) corresponding to 3:2 and 4:3 protocols), turning them into mid-gap states. 
However, exceptions occur: for example, the $\pi$-mode has minimal localization length in the 2:1 protocol.

Our analysis on the Creutz ladder shows once again that multifrequency driving is a useful tool to modify the localization length and therefore control the stability of edge modes, also in complex topological classes.

\section{Conclusions and Outlook} 
\label{sec:conclusions}

In this paper, we have described the effect of commensurate multifrequency driving on the topology of 1D systems of spinless fermions.
We have provided a generalization of the topological features observed in single-frequency drive to driving protocols with multiple commensurate frequencies.
We have first presented a general mechanism by which the topological phase boundaries are mapped to the single-frequency case.
Then we have analyzed three fermionic systems belonging to as many different symmetry classes of the Altland-Zirnbauer classification and hosting fermionic and anyonic edge modes: the Kitaev chain, the Su-Schrieffer-Heeger model, and the Creutz ladder.

While the topological phase transitions and the corresponding gap closures remain pinned at the same locations as a function of the effective driving parameters, the structure of the quasienergy spectrum greatly differs among different driving protocols.
This includes the number of edge modes in each phase and the size of the gap between edge and bulk modes.

We have shown that this variability can be harnessed to dynamically control the number and the stability of the edge modes.
Within the same driving protocol, this could be achieved for instance by (adiabatically) changing the way the intensity of the drive is distributed among different pulses, while keeping the overall period constant.
The equivalence of the effective phase diagram between many different driving protocols however introduces also a completely new handle in the way Floquet edge modes can be manipulated, namely the form of the protocol itself.
The size of the quasienergy gap and the localization length of the edge modes can be controlled by keeping the period \textit{and} the total intensity of the drive fixed, but applying different driving protocols.
We also found that zero and $\pi$-modes can be controlled independently from each other.
As a general rule of thumb, we find that more complex driving protocols tend to fill up the spectrum more evenly and reduce the size of the quasienergy gaps, thereby delocalizing the edge modes, even to the point of edge-to-edge hybridization.
However, exceptions exist where simpler protocols lead to more delocalized states.
Our work represents yet another instance of how Floquet engineering can be carefully applied to dynamically manipulate the topological properties of condensed matter systems.

In the future, to verify the control of the quasienergy gaps and of the edge modes in real time, it would be interesting to analyze the stability of the edge modes to sudden quenches between different multifrequency protocols, for instance with tensor network simulations of the fermionic chains.
Another possible direction of study consists of extending our results to other topological classes beyond two-band descriptions, for instance to the classes CII and DIII (which includes spin-orbit coupling), and to higher dimensions, including systems with anomalous edge modes~\cite{Molignini:2020}.
An extension to systems described by larger algebras would require a separate analysis to determine the conditions that allows for a simplification of the Floquet operator at the HSPs.
Finally, it would be interesting to map out the localization length of the edge modes across all different parameter regimes to provide a ``stability diagram'' of the driven systems.
This could be achieved for instance with the curvature renormalization group method~\cite{Chen:2016, Chen-Sigrist:2016, Chen:2017, Chen:2018, Chen-Schnyder:2019, Chen-Sigrist-book:2019, MoligniniReview:2019}, which is known to efficiently determine the localization length from the correlation length of the topological curvature function (\textit{e.g.} Berry curvature or Berry connection).

\acknowledgments
The author acknowledges funding from Giulio Anderheggen and the ETH Z\"{u}rich Foundation, and thanks R. Chitra and W. Chen for useful comments on the manuscript.
Computational resources from the ETH Euler cluster are gratefully acknowledged.

\appendix
%

\section{Harmonic drive in the Kitaev chain}
\label{app:harm-KC}

In the main text, we derived our results for a driving scheme consisting of a series of Dirac pulses. 
A similar quasienergy gap control can be obtained also for other types of modulations, such as a two-frequency cosine driving on top of a static field:
\begin{align}
\mu(t) &= \mu_0 + \frac{\mu_1 - \mu_0}{2}\left(  1 + \cos \left( \frac{2\pi}{T_1} t + \delta_1\right) \right) \nonumber \\
&\qquad + \frac{\mu_2 - \mu_0}{2} \left( 1 + \cos \left( \frac{2\pi}{T_2} t + \delta_2\right) \right).
\end{align}
Once again the topology is determined by the gap closure at the HSPs $k_0 = 0, \pi$. 
Evaluating the Hamiltonian at the HSPs eliminates the $\sigma^y$ part, rendering it diagonal. 
We can therefore drop the time-ordering operator in the Floquet operator, which simplifies to 
\begin{align}
U_{k_0}(T, 0) 
&= \exp \bigg[ \bigg[ -2i(\tau \cos k_0 - \mu_0) T  \nonumber \\
& \qquad  + iT \left(1 + \frac{\sin \delta_1}{2\pi p_1} \right) \Delta \mu_1 \nonumber \\
&\qquad   + iT \left(1 + \frac{\sin \delta_2}{2\pi p_2} \right) \Delta \mu_2 \bigg] \sigma^z \bigg],
\end{align}
where we have introduced $\Delta \mu_1 \equiv \mu_1 - \mu_0$, $\Delta \mu_2 \equiv \mu_2 - \mu_0$, and  expressed $T_1$ and $T_2$ in terms of $T$ and their multiplicities. 
The gap closures in the quasienergy dispersion are obtained also in this case by setting $\theta(k_0) \equiv \arccos \left( \Tr \left[ U_{k_0}(T,0) \right] /2 \right) = m \pi$, which leads to the following condition:
\begin{align}
\mu_{\text{eff}}^{\text{cos}} &\equiv \frac{\Delta \mu_1}{2} \left(1 + \frac{\sin \delta_1}{2 \pi p_1} \right)  + \frac{\Delta \mu_2}{2} \left(1 + \frac{\sin \delta_2}{2 \pi p_2} \right)  \nonumber \\
&= ( \pm \tau - \mu_0) - \frac{m \pi}{2T} 
\end{align}
For a cosine modulation, thus, the effective driving amplitude is more involved and depends explicitly on all the parameters of the driving, specifically the phase of the driving and the commensuration ratio. 
For $\delta_1 = \delta_2 = 2\pi n, n \in \mathbb{Z}$, we obtain a simpler formula similar to \eqref{condition-multifrequency-delta}, independent of the commensuration:
\begin{equation}
\frac{\Delta \mu_1 + \Delta \mu_2}{2} = ( \pm \tau - \mu_0) - \frac{m \pi}{2T}.
\end{equation}

\section{Frozen dynamics in the driven SSH model}
\label{app:frozen-SSH}

\begin{figure}[h]
\centering
\includegraphics[width=\columnwidth]{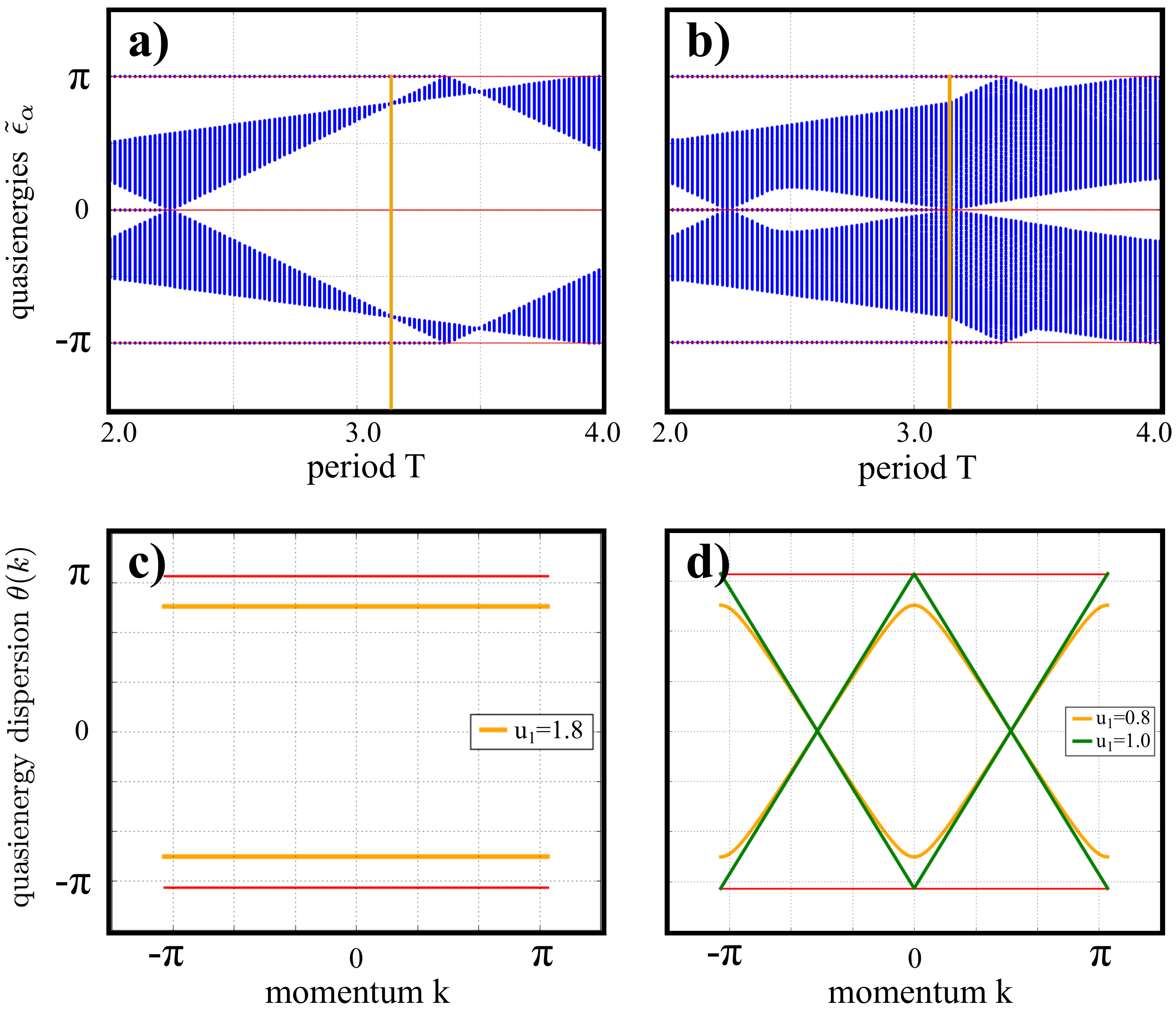}
\caption{Illustration of the effect of multifrequency driving in stabilizing edge modes via quasienergy backfolding (frozen dynamics) in the periodically driven SSH chain. Upper panels: quasienergy spectrum as a function of $T$ for a) the single-frequency drive with $v=1.0$, $u_0=0.0$, $u_1=1.8$, and b) the 2:1 double-frequency drive with $v=1.0$, $u_0=0.0$, $u_1=0.8$, $u_2=1.0$. 
The vertical orange lines depict the location of the flat band in the single-frequency case that is expanded to a fully-distributed spectrum with gap closure in the double-frequency case.
Lower panels: the corresponding quasienergy dispersions in the closed chain.
In panel d), the quasienergy dispersion at $u_1=1.0$ with coexisting gap closures at quasienergy $0$ and $\pi$ (green line) is additionally depicted.}
\label{fig:additional-gap-closures-multifrequency}
\end{figure}

In this appendix, we analyze the phenomenon of frozen dynamics emerging in the driven SSH model.
This phenomenon is analogous to what already observed in the periodically Kitaev chain~\cite{Molignini:2018} and is a consequence of the particular choice of delta-driving in which the coefficient of the Dirac is a product of intensity and period, \textit{i.e.} $u(t) = u_0 + u_1 T \sum_{m \in \mathbb{Z}} \delta(m T - t)$.
For the single-frequency drive, the quasienergy dispersion can be calculated as
\begin{align}
\theta(k) &= \arccos \bigg[ \cos(u_1 T) \cos(T E_{SSH}(k)) \nonumber \\
& \: - \frac{(u_0 + v \cos(k) ) \sin(u_1 T) \sin(T E_{SSH}(k))}{E_{SSH}(k)} \bigg]
\end{align}
with $E_{SSH}(k)$ the static energy dispersion \eqref{ssh-static-dispersion}.
We can easily see that, by setting $u_1 T = m \pi$ with $m$ an integer, the quasienergy dispersion is mapped back to the static dispersion, \textit{i.e.}
\begin{equation}
\theta(k) = \begin{cases} &T E_{SSH}(k), \qquad m \in 2\mathbb{Z}, \\
& |-\pi + T E_{SSH}(k)|, \qquad m \in 2\mathbb{Z}+1,
\end{cases}
\end{equation}
where in the second case we have restricted ourselves to $|T E_{SSH}(k) |< 2\pi$ for simplicity.
This behavior induces gap closures at non-HSPs due to the mechanism of backfolding within the Floquet-Brillouin zone explained in Ref.~\cite{Molignini:2018}.
For instance, in the case $m \in 2\mathbb{Z}$, whenever $T E_{SSH}(k)>\pi$, a gap closure occurs at $\pi$ because the quasienergy dispersion $\theta(k)$ has to be backfolded within the interval $(-\pi,\pi]$.
For every value of $E_{SSH}(k)>0$, it is always possible to find a $T$ that pushes $T E_{SSH}(k)>\pi$, and therefore these gap closures are not pinned at HSPs.
Instead, they are a particular feature of the type of delta-drive chosen here.

Besides the established frozen dynamics occurring with the single-frequency driving, the introduction of multiple Dirac pulses also creates the possibility of having frozen-dynamics-like additional gap closures at non-HSPs that are not present in the single-driving protocol.
This is best illustrated with a concrete example for the $2:1$ driving protocol at $u_0=0$, for which the Floquet operator takes the form
\begin{align}
U_{SSH,k}^{2:1}(T,0) &=  \exp \left( -i\left(  \frac{u_1 T}{2} + u_2 T \right) \left( \cos k \sigma^x + \sin k \sigma^y \right) \right)  \nonumber \\
& \quad \times \exp \left( -i \frac{T}{2} \sigma^x  \right) \nonumber \\
& \quad  \times \exp \left( -i \frac{u_1 T}{2} \left( \cos k \sigma^x + \sin k \sigma^y \right) \right) \nonumber \\
& \quad \times \exp \left( -i \frac{T}{2} \sigma^x  \right).
\label{2:1-drive-Floquet}
\end{align}
The matrix exponential can be evaluated explicitly with the help of the formula for the exponential of a Pauli vector $\boldsymbol{\sigma}$, $\exp \left( i a \hat{\mathbf{n}} \cdot \boldsymbol{\sigma} \right) = \cos a \mathds{1}_2 + i \sin a \hat{\mathbf{n}} \cdot \boldsymbol{\sigma}$, where $\hat{\mathbf{n}}$ is a unit vector, such as $(\cos k, \sin k, 0)$ in our case.
By inserting this formula in Eq.~\ref{2:1-drive-Floquet}, we can derive
the dispersion
\begin{widetext}
\begin{align}
\theta(k) 
&=  \arccos \bigg[ 2 \cos^2 \left(T/2 \right) \cos \big( (u_1+u_2)T \big) + 2 \sin^2 (T/2) \left( - \cos(u_1 T/2) \cos\big( (u_1 + 2u_2)T/2 \big) \right. \nonumber \\
& \left.  \qquad \qquad\qquad + \cos(2k) \sin(u_1T/2) \sin \big( (u_1 + 2u_2)T/2 \big) \right) - 2 \cos k \sin T \sin((u_1 + u_2)T) \big].
\end{align}
\end{widetext}
This dispersion acquires a very simple form for multiples of $u_2T=\pi$, namely
\begin{equation}
\theta(k) = \arccos  \left[ 2 \left(  1 - \sin^2 \left( \frac{\pi u_1}{2} \right)  \left( 1 + \cos(2k) \right) \right) \right]
\end{equation}
which exhibits linear gap closures at $k = \pm \frac{\pi}{2}$.
An example of these additional gap closures and the corresponding edge states is shown in Fig.~\ref{fig:additional-gap-closures-multifrequency}, where the quasienergy spectrum (OBC) and the quasienergy dispersion (PBC) of the single-frequency and of the 2:1 double frequency protocol are compared.
It is in particular interesting to note how the flat band in the single-frequency case is converted into a gap closing dispersion with the double-frequency drive.
As we can gauge from the figure, this procedure stabilizes the edge state at quasienergy $\tilde{\epsilon}=0$, which survives for longer periods until it gets annihilated when the gap closes at $T=\pi$.
Note also that the coexistence of both types of frozen dynamics (\textit{i.e.} $u_1T=m\pi$ and $u_2T=\tilde{m} \pi$), implies $\theta = \arccos[-2\cos(2k)]$, which has both a $0$-gap closure at $k= \pm \frac{\pi}{2}$ and a $\pi$-gap closure at $k=0, \pm \pi$ (Fig.~\ref{fig:additional-gap-closures-multifrequency}d)).


\section{Frozen dynamics in the driven Creutz ladder}
\label{app:frozen-CL}

\begin{figure}
\centering
\includegraphics[width=0.9\columnwidth]{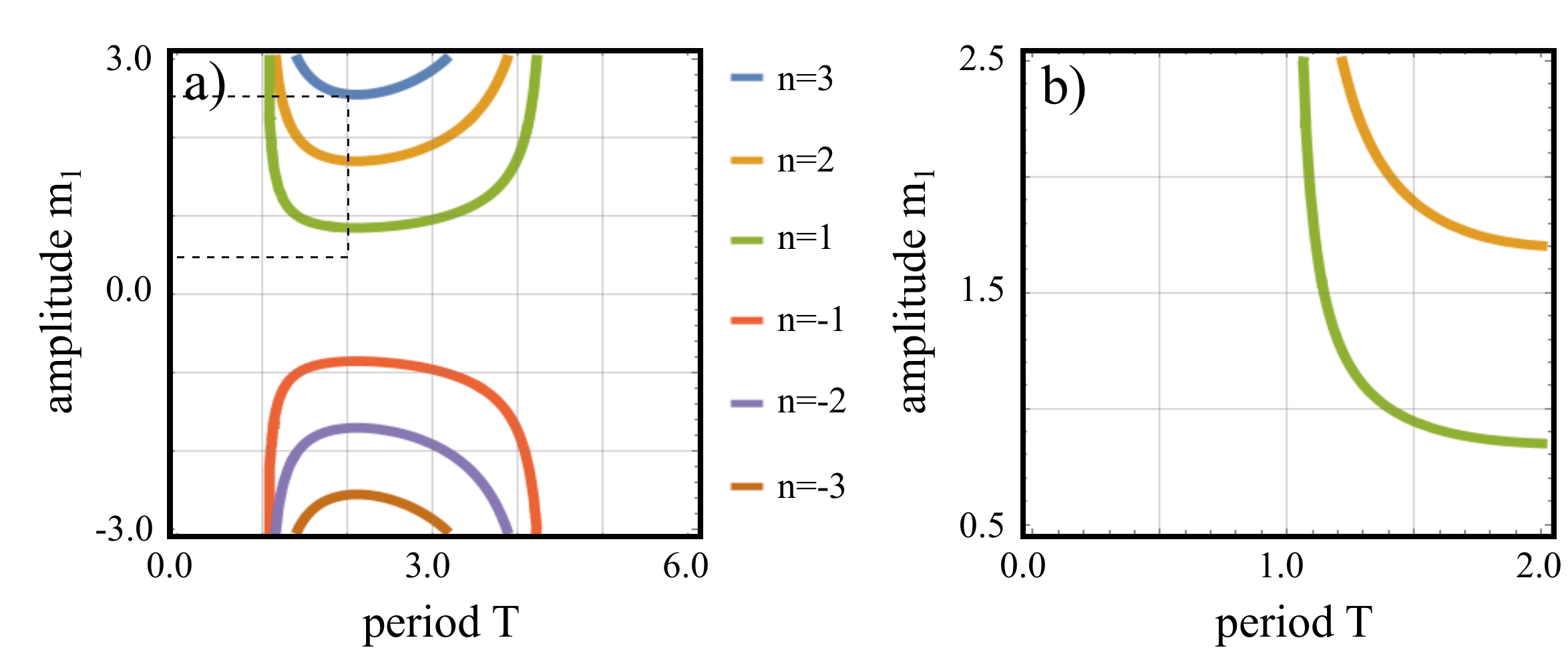}
\caption{a) Lines of frozen dynamics in the driven Creutz ladder corresponding to various values of $n$ in \eqref{frozen-dynamics-CL-n}. b) zoom of panel a) to fit the boundaries of the phase diagrams depicted in Fig.~\ref{fig:PD-creutz}.}
\label{fig:frozen-dynamics-CL}
\end{figure}

In this appendix, we show how frozen dynamics arises in the driven Creutz ladder.
For the sake of exposition, we focus on the single-frequency case with $\theta= \pi/2$ and $m_0=1.0$, but these considerations are independent of the choice of parameters.
For this driving protocol, the quasienergy dispersion is calculated as
\begin{widetext}
\begin{align}
\theta(k) &= \arccos \bigg[ \cos \left( S(k) \right)  \cos \left( T E_{CL}(k) \right) - \frac{\left( 1 + 4 \cos^2(k) - 2\sin(k) \right)}{\sqrt{3 + 2 \cos(2k)} E_{CL}(k)}  \sin \left( S(k)  \right)  \sin \left( T  E_{CL}(k) \right)  \bigg],
\end{align}
\end{widetext}
where $E_{CL}(k) \equiv \sqrt{4 \cos^2(k) + (1 - 2\sin(k))^2}$ is the positive branch of the static dispersion (for $\theta= \pi/2$ and $m_0 = m =1.0$) and $S(k) \equiv m_1 T \sqrt{3+2\cos(2k)}$.
By examining the shape of the dispersion, we can see that we can obtain frozen dynamics whenever $ m_1 T \sqrt{3+2 \cos(2k)}=n \pi$, with $n \in \mathbb{Z}$.
Interestingly enough, in this case the corresponding transition lines have an additional $k$-dependence. 
This can however be eliminated self-consistently by demanding that $\theta(k) = 0, \pm \pi$.
The only real solution for $k$ is $k= \arcsin \left( \frac{5}{4} - \frac{\pi^2}{4T^2} \right)$. 
Inserting this in the equation for the frozen dynamics, we obtain the explicit equation
\begin{equation}
m_1(T) = \frac{n \pi}{T \sqrt{3+2 \cos \left(2  \arcsin \left( \frac{5}{4} - \frac{\pi^2}{4T^2} \right) \right)}}.
\label{frozen-dynamics-CL-n}
\end{equation}
This equation is plotted in Fig.~\ref{fig:frozen-dynamics-CL} for various values of $n$. 
Note that only the lines for $n=1,2$ are visible in the region where the phase diagrams of Fig.~\ref{fig:PD-creutz} have been plotted.

\begin{figure}
\centering
\includegraphics[width=0.9\columnwidth]{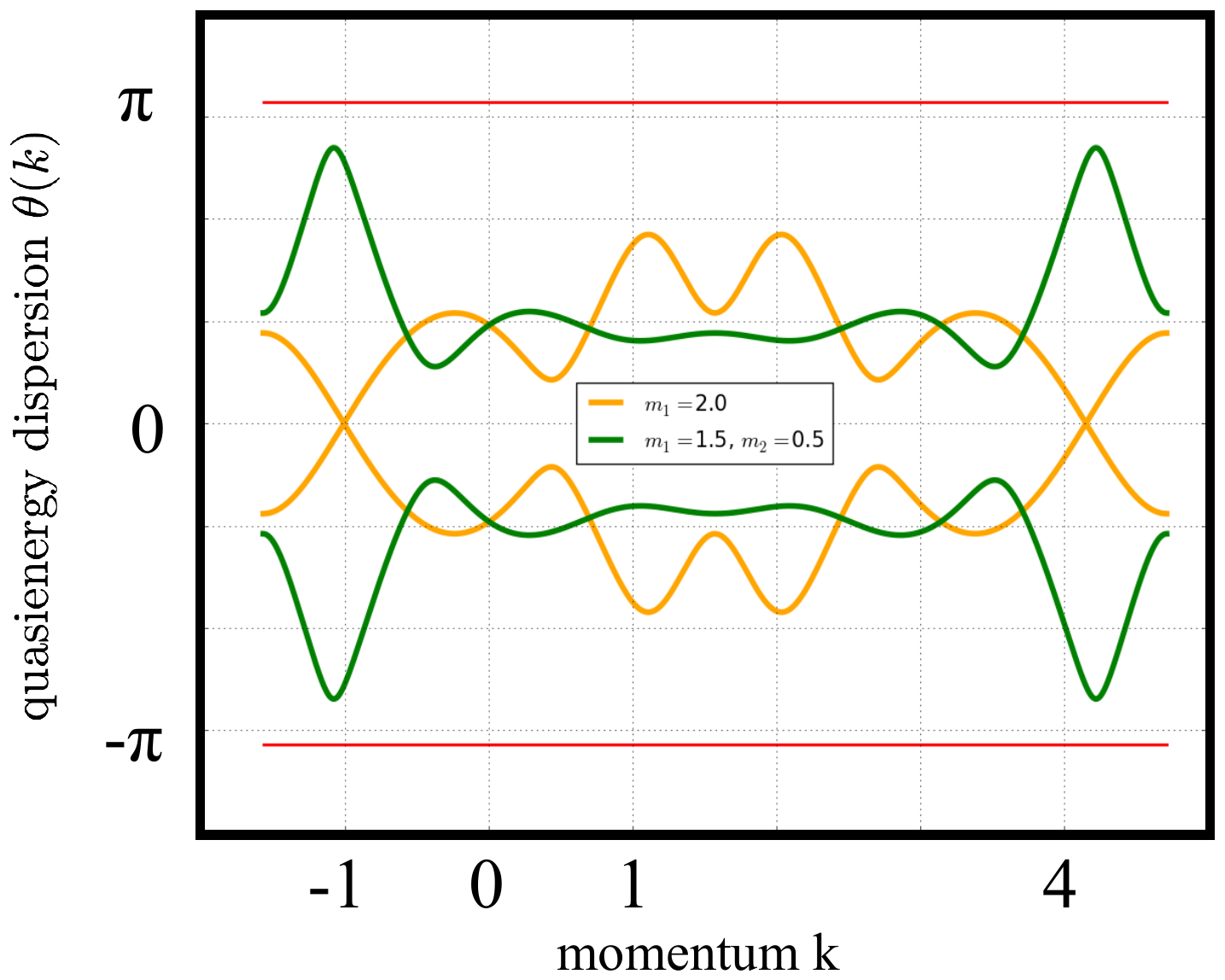}
\caption{Comparison of the quasienergy dispersion for the driven Creutz ladder with the single-driving (orange curve) and double-driving (green curve) protocols at $T=1.08$, corresponding to the gap closure marked by the red arrow in Fig.~\ref{fig:PD-creutz}. This gap closure does not occur at a HSP, and is lifted in the 2:1 driving protocol. The corresponding transition is due to the specific form of the delta-drive.}
\label{fig:comparison-disp-creutz}
\end{figure}

The line for $n=1$ indeed corresponds to an additional gap closure in the quasienergy spectrum where the number of zero modes changes (cf. red arrow of Fig.~\ref{fig:PD-creutz} and Fig.~\ref{fig:multifrequency-quasienergies-comparison-creutz}a)).
A closer inspection of the quasienergy \textit{dispersion} at this point, shown in Fig.~\ref{fig:comparison-disp-creutz} reveals indeed that the gap closure occurs at a non-HSP $k\neq 0, \pm \pi/2, \pm \pi$.
For multifrequency driving protocols, this transitions disappears.
The line for $n=2$, on the other hand, does not seem to be systematically associated with a TPT in the phase diagram.
This could be due to the fact that the number of edge modes does not systematically change despite the gap closure.

Altogether, these results seem to indicate that frozen dynamics is quite erratic and does not represent a reliable and systematic way to obtain robust TPTs.
A methodical analysis of frozen dynamics is left for future studies.

\bibliography{many-body-biblio}

\begin{thebibliography}{98}
\expandafter\ifx\csname natexlab\endcsname\relax\def\natexlab#1{#1}\fi
\expandafter\ifx\csname bibnamefont\endcsname\relax
  \def\bibnamefont#1{#1}\fi
\expandafter\ifx\csname bibfnamefont\endcsname\relax
  \def\bibfnamefont#1{#1}\fi
\expandafter\ifx\csname citenamefont\endcsname\relax
  \def\citenamefont#1{#1}\fi
\expandafter\ifx\csname url\endcsname\relax
  \def\url#1{\texttt{#1}}\fi
\expandafter\ifx\csname urlprefix\endcsname\relax\def\urlprefix{URL }\fi
\providecommand{\bibinfo}[2]{#2}
\providecommand{\eprint}[2][]{\url{#2}}

\bibitem[{\citenamefont{Landau}(1937)}]{Landau}
\bibinfo{author}{\bibfnamefont{L.~D.} \bibnamefont{Landau}},
  \bibinfo{journal}{Zh. Eksp. Teor. Fiz.} \textbf{\bibinfo{volume}{7}},
  \bibinfo{pages}{19} (\bibinfo{year}{1937}).

\bibitem[{\citenamefont{Miransky}(1994)}]{Miransky-book}
\bibinfo{author}{\bibfnamefont{V.~A.} \bibnamefont{Miransky}},
  \emph{\bibinfo{title}{Dynamical Symmetry Breaking in Quantum Field Theories}}
  (\bibinfo{publisher}{World Scientific Publishing Co.}, \bibinfo{year}{1994}).

\bibitem[{\citenamefont{Wen}(1990)}]{Wen:1990}
\bibinfo{author}{\bibfnamefont{X.-G.} \bibnamefont{Wen}},
  \bibinfo{journal}{Int. J. Mod. Phys. B} \textbf{\bibinfo{volume}{4}},
  \bibinfo{pages}{239} (\bibinfo{year}{1990}).

\bibitem[{\citenamefont{Thouless et~al.}(1982)\citenamefont{Thouless, Kohmoto,
  Nightingale, and den Nijs}}]{Thouless:1982}
\bibinfo{author}{\bibfnamefont{D.~J.} \bibnamefont{Thouless}},
  \bibinfo{author}{\bibfnamefont{M.}~\bibnamefont{Kohmoto}},
  \bibinfo{author}{\bibfnamefont{M.~P.} \bibnamefont{Nightingale}},
  \bibnamefont{and} \bibinfo{author}{\bibfnamefont{M.}~\bibnamefont{den Nijs}},
  \bibinfo{journal}{Phys. Rev. Lett.} \textbf{\bibinfo{volume}{49}},
  \bibinfo{pages}{405} (\bibinfo{year}{1982}).

\bibitem[{\citenamefont{Wen}(1989)}]{Wen:1989}
\bibinfo{author}{\bibfnamefont{X.-G.} \bibnamefont{Wen}},
  \bibinfo{journal}{Phys. Rev. B} \textbf{\bibinfo{volume}{40}},
  \bibinfo{pages}{7387} (\bibinfo{year}{1989}).

\bibitem[{\citenamefont{Qi et~al.}(2008)\citenamefont{Qi, Hughes, and
  Zhang}}]{Qi:2008}
\bibinfo{author}{\bibfnamefont{X.-L.} \bibnamefont{Qi}},
  \bibinfo{author}{\bibfnamefont{T.}~\bibnamefont{Hughes}}, \bibnamefont{and}
  \bibinfo{author}{\bibfnamefont{S.-C.} \bibnamefont{Zhang}},
  \bibinfo{journal}{Phys. Rev. B} \textbf{\bibinfo{volume}{78}},
  \bibinfo{pages}{195424} (\bibinfo{year}{2008}).

\bibitem[{\citenamefont{Goldman and Su}(1995)}]{Goldman:1995}
\bibinfo{author}{\bibfnamefont{V.~J.} \bibnamefont{Goldman}} \bibnamefont{and}
  \bibinfo{author}{\bibfnamefont{B.}~\bibnamefont{Su}},
  \bibinfo{journal}{Science} \textbf{\bibinfo{volume}{267 (5200)}},
  \bibinfo{pages}{1010} (\bibinfo{year}{1995}).

\bibitem[{\citenamefont{de~Picciotto et~al.}(1997)\citenamefont{de~Picciotto,
  Reznikov, Heiblum, Umansky, Bunin, and Mahalu}}]{dePicciotto:1997}
\bibinfo{author}{\bibfnamefont{R.}~\bibnamefont{de~Picciotto}},
  \bibinfo{author}{\bibfnamefont{M.}~\bibnamefont{Reznikov}},
  \bibinfo{author}{\bibfnamefont{M.}~\bibnamefont{Heiblum}},
  \bibinfo{author}{\bibfnamefont{V.}~\bibnamefont{Umansky}},
  \bibinfo{author}{\bibfnamefont{G.}~\bibnamefont{Bunin}}, \bibnamefont{and}
  \bibinfo{author}{\bibfnamefont{D.}~\bibnamefont{Mahalu}},
  \bibinfo{journal}{Nature} \textbf{\bibinfo{volume}{389 (6647)}},
  \bibinfo{pages}{162} (\bibinfo{year}{1997}).

\bibitem[{\citenamefont{Martin et~al.}(2004)\citenamefont{Martin, Ilani,
  Verdene, Smet, Umansky, Mahalu, Schuh, Abstreiter, and Yacoby}}]{Martin:2004}
\bibinfo{author}{\bibfnamefont{J.}~\bibnamefont{Martin}},
  \bibinfo{author}{\bibfnamefont{S.}~\bibnamefont{Ilani}},
  \bibinfo{author}{\bibfnamefont{B.}~\bibnamefont{Verdene}},
  \bibinfo{author}{\bibfnamefont{J.}~\bibnamefont{Smet}},
  \bibinfo{author}{\bibfnamefont{V.}~\bibnamefont{Umansky}},
  \bibinfo{author}{\bibfnamefont{D.}~\bibnamefont{Mahalu}},
  \bibinfo{author}{\bibfnamefont{D.}~\bibnamefont{Schuh}},
  \bibinfo{author}{\bibfnamefont{G.}~\bibnamefont{Abstreiter}},
  \bibnamefont{and} \bibinfo{author}{\bibfnamefont{A.}~\bibnamefont{Yacoby}},
  \bibinfo{journal}{Science} \textbf{\bibinfo{volume}{305 (5686)}},
  \bibinfo{pages}{980} (\bibinfo{year}{2004}).

\bibitem[{\citenamefont{Moore and Read}(1991)}]{Moore}
\bibinfo{author}{\bibfnamefont{G.}~\bibnamefont{Moore}} \bibnamefont{and}
  \bibinfo{author}{\bibfnamefont{N.}~\bibnamefont{Read}},
  \bibinfo{journal}{Nucl. Phys.} \textbf{\bibinfo{volume}{B360}},
  \bibinfo{pages}{362} (\bibinfo{year}{1991}).

\bibitem[{\citenamefont{Kitaev}(2001)}]{Kitaev:2001}
\bibinfo{author}{\bibfnamefont{A.~Y.} \bibnamefont{Kitaev}},
  \bibinfo{journal}{Phys.-Usp.} \textbf{\bibinfo{volume}{44}},
  \bibinfo{pages}{131} (\bibinfo{year}{2001}).

\bibitem[{\citenamefont{Mong et~al.}(2014)\citenamefont{Mong, Clarke, Alicea,
  Lindner, Fendley, Nayak, Oreg, Stern, Berg, Shtengel et~al.}}]{Mong:2014}
\bibinfo{author}{\bibfnamefont{R.~S.~K.} \bibnamefont{Mong}},
  \bibinfo{author}{\bibfnamefont{D.~J.} \bibnamefont{Clarke}},
  \bibinfo{author}{\bibfnamefont{J.}~\bibnamefont{Alicea}},
  \bibinfo{author}{\bibfnamefont{N.~H.} \bibnamefont{Lindner}},
  \bibinfo{author}{\bibfnamefont{P.}~\bibnamefont{Fendley}},
  \bibinfo{author}{\bibfnamefont{C.}~\bibnamefont{Nayak}},
  \bibinfo{author}{\bibfnamefont{Y.}~\bibnamefont{Oreg}},
  \bibinfo{author}{\bibfnamefont{A.}~\bibnamefont{Stern}},
  \bibinfo{author}{\bibfnamefont{E.}~\bibnamefont{Berg}},
  \bibinfo{author}{\bibfnamefont{K.}~\bibnamefont{Shtengel}},
  \bibnamefont{et~al.}, \bibinfo{journal}{Phys. Rev. X}
  \textbf{\bibinfo{volume}{4}}, \bibinfo{pages}{011036} (\bibinfo{year}{2014}).

\bibitem[{\citenamefont{Wen}(1995)}]{Wen:1995}
\bibinfo{author}{\bibfnamefont{X.-G.} \bibnamefont{Wen}},
  \bibinfo{journal}{Advances in Physics} \textbf{\bibinfo{volume}{44}},
  \bibinfo{pages}{405} (\bibinfo{year}{1995}).

\bibitem[{\citenamefont{Chen et~al.}(2010)\citenamefont{Chen, Gu, and
  Wen}}]{XieChen:2010}
\bibinfo{author}{\bibfnamefont{X.}~\bibnamefont{Chen}},
  \bibinfo{author}{\bibfnamefont{Z.-C.} \bibnamefont{Gu}}, \bibnamefont{and}
  \bibinfo{author}{\bibfnamefont{X.-G.} \bibnamefont{Wen}},
  \bibinfo{journal}{Phys. Rev. B} \textbf{\bibinfo{volume}{82}},
  \bibinfo{pages}{155138} (\bibinfo{year}{2010}).

\bibitem[{\citenamefont{Fidkowski}(2010)}]{Fidkowski:2010}
\bibinfo{author}{\bibfnamefont{L.}~\bibnamefont{Fidkowski}},
  \bibinfo{journal}{Phys. Rev. Lett.} \textbf{\bibinfo{volume}{104}},
  \bibinfo{pages}{130502} (\bibinfo{year}{2010}).

\bibitem[{\citenamefont{Kitaev}(2003)}]{Kitaev:2003}
\bibinfo{author}{\bibfnamefont{A.~Y.} \bibnamefont{Kitaev}},
  \bibinfo{journal}{Ann. Phys. (N.Y.)} \textbf{\bibinfo{volume}{303}},
  \bibinfo{pages}{2} (\bibinfo{year}{2003}).

\bibitem[{\citenamefont{Nayak et~al.}(2008)\citenamefont{Nayak, Simon, Stern,
  Freedman, and Sarma}}]{Nayak:2008}
\bibinfo{author}{\bibfnamefont{C.}~\bibnamefont{Nayak}},
  \bibinfo{author}{\bibfnamefont{S.~H.} \bibnamefont{Simon}},
  \bibinfo{author}{\bibfnamefont{A.}~\bibnamefont{Stern}},
  \bibinfo{author}{\bibfnamefont{M.}~\bibnamefont{Freedman}}, \bibnamefont{and}
  \bibinfo{author}{\bibfnamefont{S.~D.} \bibnamefont{Sarma}},
  \bibinfo{journal}{Rev. Mod. Phys.} \textbf{\bibinfo{volume}{80}},
  \bibinfo{pages}{1083} (\bibinfo{year}{2008}).

\bibitem[{\citenamefont{Vandenberge and Fischetti}(2017)}]{Vandenberghe:2017}
\bibinfo{author}{\bibfnamefont{W.~G.} \bibnamefont{Vandenberge}}
  \bibnamefont{and} \bibinfo{author}{\bibfnamefont{M.~V.}
  \bibnamefont{Fischetti}}, \bibinfo{journal}{Nat. Comm.}
  \textbf{\bibinfo{volume}{8}}, \bibinfo{pages}{14184} (\bibinfo{year}{2017}).

\bibitem[{\citenamefont{Wang et~al.}(2016)\citenamefont{Wang, Lang, and
  Kou}}]{WangBook:2016}
\bibinfo{author}{\bibfnamefont{K.~L.} \bibnamefont{Wang}},
  \bibinfo{author}{\bibfnamefont{M.}~\bibnamefont{Lang}}, \bibnamefont{and}
  \bibinfo{author}{\bibfnamefont{X.}~\bibnamefont{Kou}},
  \emph{\bibinfo{title}{Spintronics of Topological Insulators. In: Handbook of
  Spintronics}} (\bibinfo{publisher}{Springer}, \bibinfo{address}{Dordrecht},
  \bibinfo{year}{2016}).

\bibitem[{\citenamefont{Kitagawa et~al.}(2010)\citenamefont{Kitagawa, Berg,
  Rudner, and Demler}}]{Kitagawa:2010}
\bibinfo{author}{\bibfnamefont{T.}~\bibnamefont{Kitagawa}},
  \bibinfo{author}{\bibfnamefont{E.}~\bibnamefont{Berg}},
  \bibinfo{author}{\bibfnamefont{M.}~\bibnamefont{Rudner}}, \bibnamefont{and}
  \bibinfo{author}{\bibfnamefont{E.}~\bibnamefont{Demler}},
  \bibinfo{journal}{Phys. Rev. B} \textbf{\bibinfo{volume}{82}},
  \bibinfo{pages}{235114} (\bibinfo{year}{2010}).

\bibitem[{\citenamefont{Lindner et~al.}(2011)\citenamefont{Lindner, Refael, and
  Galitski}}]{Lindner:2011}
\bibinfo{author}{\bibfnamefont{N.~H.} \bibnamefont{Lindner}},
  \bibinfo{author}{\bibfnamefont{G.}~\bibnamefont{Refael}}, \bibnamefont{and}
  \bibinfo{author}{\bibfnamefont{V.}~\bibnamefont{Galitski}},
  \bibinfo{journal}{Nat. Phys.} \textbf{\bibinfo{volume}{7}},
  \bibinfo{pages}{490} (\bibinfo{year}{2011}).

\bibitem[{\citenamefont{Cayssol et~al.}(2013)\citenamefont{Cayssol, D\'{o}ra,
  Simon, and Moessner}}]{Cayssol:2013}
\bibinfo{author}{\bibfnamefont{J.}~\bibnamefont{Cayssol}},
  \bibinfo{author}{\bibfnamefont{B.}~\bibnamefont{D\'{o}ra}},
  \bibinfo{author}{\bibfnamefont{F.}~\bibnamefont{Simon}}, \bibnamefont{and}
  \bibinfo{author}{\bibfnamefont{R.}~\bibnamefont{Moessner}},
  \bibinfo{journal}{Phys. Status Solidi RRL} \textbf{\bibinfo{volume}{7}},
  \bibinfo{pages}{101} (\bibinfo{year}{2013}).

\bibitem[{\citenamefont{Harper and Roy}(2017)}]{Harper:2017}
\bibinfo{author}{\bibfnamefont{F.}~\bibnamefont{Harper}} \bibnamefont{and}
  \bibinfo{author}{\bibfnamefont{R.}~\bibnamefont{Roy}},
  \bibinfo{journal}{Phys. Rev. Lett.} \textbf{\bibinfo{volume}{118}},
  \bibinfo{pages}{115301} (\bibinfo{year}{2017}).

\bibitem[{\citenamefont{Roy and Harper}(2017)}]{Roy:2017}
\bibinfo{author}{\bibfnamefont{R.}~\bibnamefont{Roy}} \bibnamefont{and}
  \bibinfo{author}{\bibfnamefont{F.}~\bibnamefont{Harper}},
  \bibinfo{journal}{Phys. Rev. B} \textbf{\bibinfo{volume}{96}},
  \bibinfo{pages}{155118} (\bibinfo{year}{2017}).

\bibitem[{\citenamefont{Esin et~al.}(2018)\citenamefont{Esin, Rudner, Refael,
  and Lindner}}]{Esin:2018}
\bibinfo{author}{\bibfnamefont{I.}~\bibnamefont{Esin}},
  \bibinfo{author}{\bibfnamefont{M.~S.} \bibnamefont{Rudner}},
  \bibinfo{author}{\bibfnamefont{G.}~\bibnamefont{Refael}}, \bibnamefont{and}
  \bibinfo{author}{\bibfnamefont{N.~H.} \bibnamefont{Lindner}},
  \bibinfo{journal}{Phys. Rev. B} \textbf{\bibinfo{volume}{97}},
  \bibinfo{pages}{245401} (\bibinfo{year}{2018}).

\bibitem[{\citenamefont{Liu et~al.}(2013)\citenamefont{Liu, Levchenko, and
  Baranger}}]{Liu}
\bibinfo{author}{\bibfnamefont{D.~E.} \bibnamefont{Liu}},
  \bibinfo{author}{\bibfnamefont{A.}~\bibnamefont{Levchenko}},
  \bibnamefont{and} \bibinfo{author}{\bibfnamefont{H.~U.}
  \bibnamefont{Baranger}}, \bibinfo{journal}{Phys. Rev. Lett.}
  \textbf{\bibinfo{volume}{111}}, \bibinfo{pages}{047002}
  (\bibinfo{year}{2013}).

\bibitem[{\citenamefont{Thakurathi et~al.}(2013)\citenamefont{Thakurathi,
  Patel, Sen, and Dutta}}]{Thakurathi:2013}
\bibinfo{author}{\bibfnamefont{M.}~\bibnamefont{Thakurathi}},
  \bibinfo{author}{\bibfnamefont{A.~A.} \bibnamefont{Patel}},
  \bibinfo{author}{\bibfnamefont{D.}~\bibnamefont{Sen}}, \bibnamefont{and}
  \bibinfo{author}{\bibfnamefont{A.}~\bibnamefont{Dutta}},
  \bibinfo{journal}{Phys. Rev. B} \textbf{\bibinfo{volume}{88}},
  \bibinfo{pages}{155133} (\bibinfo{year}{2013}).

\bibitem[{\citenamefont{Benito et~al.}(2014)\citenamefont{Benito,
  G\'{o}mez-Le\'{o}n, Bastidas, Brandes, and Platero}}]{Benito:2014}
\bibinfo{author}{\bibfnamefont{M.}~\bibnamefont{Benito}},
  \bibinfo{author}{\bibfnamefont{A.}~\bibnamefont{G\'{o}mez-Le\'{o}n}},
  \bibinfo{author}{\bibfnamefont{V.~M.} \bibnamefont{Bastidas}},
  \bibinfo{author}{\bibfnamefont{T.}~\bibnamefont{Brandes}}, \bibnamefont{and}
  \bibinfo{author}{\bibfnamefont{G.}~\bibnamefont{Platero}},
  \bibinfo{journal}{Phys. Rev. B} \textbf{\bibinfo{volume}{90}},
  \bibinfo{pages}{205127} (\bibinfo{year}{2014}).

\bibitem[{\citenamefont{Thakurathi et~al.}(2014)\citenamefont{Thakurathi,
  Sengupta, and Sen}}]{Thakurathi:2014}
\bibinfo{author}{\bibfnamefont{M.}~\bibnamefont{Thakurathi}},
  \bibinfo{author}{\bibfnamefont{K.}~\bibnamefont{Sengupta}}, \bibnamefont{and}
  \bibinfo{author}{\bibfnamefont{D.}~\bibnamefont{Sen}},
  \bibinfo{journal}{Phys. Rev. B} \textbf{\bibinfo{volume}{89}},
  \bibinfo{pages}{235434} (\bibinfo{year}{2014}).

\bibitem[{\citenamefont{Wang et~al.}(2014{\natexlab{a}})\citenamefont{Wang,
  Q.-F.-Sun, and Xie}}]{Wang:2014}
\bibinfo{author}{\bibfnamefont{P.}~\bibnamefont{Wang}},
  \bibinfo{author}{\bibnamefont{Q.-F.-Sun}}, \bibnamefont{and}
  \bibinfo{author}{\bibfnamefont{X.~C.} \bibnamefont{Xie}},
  \bibinfo{journal}{Phys. Rev. B} \textbf{\bibinfo{volume}{90}},
  \bibinfo{pages}{155407} (\bibinfo{year}{2014}{\natexlab{a}}).

\bibitem[{\citenamefont{Sacramento}(2015)}]{Sacramento:2015}
\bibinfo{author}{\bibfnamefont{P.~D.} \bibnamefont{Sacramento}},
  \bibinfo{journal}{Phys. Rev. B} \textbf{\bibinfo{volume}{91}},
  \bibinfo{pages}{214518} (\bibinfo{year}{2015}).

\bibitem[{\citenamefont{Thakurathi et~al.}(2017)\citenamefont{Thakurathi, Loss,
  and Klinovaja}}]{Thakurathi:2017}
\bibinfo{author}{\bibfnamefont{M.}~\bibnamefont{Thakurathi}},
  \bibinfo{author}{\bibfnamefont{D.}~\bibnamefont{Loss}}, \bibnamefont{and}
  \bibinfo{author}{\bibfnamefont{J.}~\bibnamefont{Klinovaja}},
  \bibinfo{journal}{Phys. Rev. B} \textbf{\bibinfo{volume}{95}},
  \bibinfo{pages}{155407} (\bibinfo{year}{2017}).

\bibitem[{\citenamefont{Molignini et~al.}(2017)\citenamefont{Molignini, van
  Nieuwenburg, and Chitra}}]{Molignini:2017}
\bibinfo{author}{\bibfnamefont{P.}~\bibnamefont{Molignini}},
  \bibinfo{author}{\bibfnamefont{E.}~\bibnamefont{van Nieuwenburg}},
  \bibnamefont{and} \bibinfo{author}{\bibfnamefont{R.}~\bibnamefont{Chitra}},
  \bibinfo{journal}{Phys. Rev. B} \textbf{\bibinfo{volume}{96}},
  \bibinfo{pages}{125144} (\bibinfo{year}{2017}).

\bibitem[{\citenamefont{Molignini et~al.}(2018)\citenamefont{Molignini, Chen,
  and Chitra}}]{Molignini:2018}
\bibinfo{author}{\bibfnamefont{P.}~\bibnamefont{Molignini}},
  \bibinfo{author}{\bibfnamefont{W.}~\bibnamefont{Chen}}, \bibnamefont{and}
  \bibinfo{author}{\bibfnamefont{R.}~\bibnamefont{Chitra}},
  \bibinfo{journal}{Phys. Rev. B} \textbf{\bibinfo{volume}{98}},
  \bibinfo{pages}{125129} (\bibinfo{year}{2018}).

\bibitem[{\citenamefont{{\v C}ade{\v z} et~al.}(2019)\citenamefont{{\v C}ade{\v
  z}, Mondaini, and Sacramento}}]{Cadez:2019}
\bibinfo{author}{\bibfnamefont{T.}~\bibnamefont{{\v C}ade{\v z}}},
  \bibinfo{author}{\bibfnamefont{R.}~\bibnamefont{Mondaini}}, \bibnamefont{and}
  \bibinfo{author}{\bibfnamefont{P.~D.} \bibnamefont{Sacramento}},
  \bibinfo{journal}{Phys. Rev. B} \textbf{\bibinfo{volume}{99}},
  \bibinfo{pages}{014301} (\bibinfo{year}{2019}).

\bibitem[{\citenamefont{Wang et~al.}(2014{\natexlab{b}})\citenamefont{Wang,
  Wang, Shen, L.Sheng, Xing, and Savrasov}}]{Wang:2014-Weyl-semimetal}
\bibinfo{author}{\bibfnamefont{R.}~\bibnamefont{Wang}},
  \bibinfo{author}{\bibfnamefont{B.}~\bibnamefont{Wang}},
  \bibinfo{author}{\bibfnamefont{R.}~\bibnamefont{Shen}},
  \bibinfo{author}{\bibnamefont{L.Sheng}},
  \bibinfo{author}{\bibfnamefont{D.}~\bibnamefont{Xing}}, \bibnamefont{and}
  \bibinfo{author}{\bibfnamefont{S.~Y.} \bibnamefont{Savrasov}},
  \bibinfo{journal}{Europhys. Lett.} \textbf{\bibinfo{volume}{105}},
  \bibinfo{pages}{17004} (\bibinfo{year}{2014}{\natexlab{b}}).

\bibitem[{\citenamefont{Chan et~al.}(2016)\citenamefont{Chan, Lee, Burch, Han,
  and Ran}}]{Chan:2016-Weyl}
\bibinfo{author}{\bibfnamefont{C.-K.} \bibnamefont{Chan}},
  \bibinfo{author}{\bibfnamefont{P.~A.} \bibnamefont{Lee}},
  \bibinfo{author}{\bibfnamefont{K.~S.} \bibnamefont{Burch}},
  \bibinfo{author}{\bibfnamefont{J.~H.} \bibnamefont{Han}}, \bibnamefont{and}
  \bibinfo{author}{\bibfnamefont{Y.}~\bibnamefont{Ran}}, \bibinfo{journal}{Phys
  Rev. Lett.} \textbf{\bibinfo{volume}{116}}, \bibinfo{pages}{026805}
  (\bibinfo{year}{2016}).

\bibitem[{\citenamefont{Yan and Wang}(2016)}]{Yan:2016}
\bibinfo{author}{\bibfnamefont{Z.}~\bibnamefont{Yan}} \bibnamefont{and}
  \bibinfo{author}{\bibfnamefont{Z.}~\bibnamefont{Wang}},
  \bibinfo{journal}{Phys Rev. Lett.} \textbf{\bibinfo{volume}{117}},
  \bibinfo{pages}{087402} (\bibinfo{year}{2016}).

\bibitem[{\citenamefont{Bucciantini et~al.}(2017)\citenamefont{Bucciantini,
  Roy, Kitamura, and Oka}}]{Bucciantini:2017}
\bibinfo{author}{\bibfnamefont{L.}~\bibnamefont{Bucciantini}},
  \bibinfo{author}{\bibfnamefont{S.}~\bibnamefont{Roy}},
  \bibinfo{author}{\bibfnamefont{S.}~\bibnamefont{Kitamura}}, \bibnamefont{and}
  \bibinfo{author}{\bibfnamefont{T.}~\bibnamefont{Oka}},
  \bibinfo{journal}{Phys. Rev. B} \textbf{\bibinfo{volume}{96}},
  \bibinfo{pages}{041126(R)} (\bibinfo{year}{2017}).

\bibitem[{\citenamefont{H{\"u}bener et~al.}(2017)\citenamefont{H{\"u}bener,
  Sentef, Giovannini, Kemper, and Rubio}}]{Huebener:2017}
\bibinfo{author}{\bibfnamefont{H.}~\bibnamefont{H{\"u}bener}},
  \bibinfo{author}{\bibfnamefont{M.~A.} \bibnamefont{Sentef}},
  \bibinfo{author}{\bibfnamefont{U.~D.} \bibnamefont{Giovannini}},
  \bibinfo{author}{\bibfnamefont{A.~F.} \bibnamefont{Kemper}},
  \bibnamefont{and} \bibinfo{author}{\bibfnamefont{A.}~\bibnamefont{Rubio}},
  \bibinfo{journal}{Nat. Comm.} \textbf{\bibinfo{volume}{8}},
  \bibinfo{pages}{13940} (\bibinfo{year}{2017}).

\bibitem[{\citenamefont{Cao et~al.}(2017)\citenamefont{Cao, Qi, and
  Xiang}}]{Cao:2017}
\bibinfo{author}{\bibfnamefont{J.}~\bibnamefont{Cao}},
  \bibinfo{author}{\bibfnamefont{F.}~\bibnamefont{Qi}}, \bibnamefont{and}
  \bibinfo{author}{\bibfnamefont{Y.}~\bibnamefont{Xiang}},
  \bibinfo{journal}{Europhys. Lett.} \textbf{\bibinfo{volume}{119}},
  \bibinfo{pages}{57008} (\bibinfo{year}{2017}).

\bibitem[{\citenamefont{Chen et~al.}(2018)\citenamefont{Chen, Zhou, and
  Xu}}]{ChenZhou:2018}
\bibinfo{author}{\bibfnamefont{R.}~\bibnamefont{Chen}},
  \bibinfo{author}{\bibfnamefont{B.}~\bibnamefont{Zhou}}, \bibnamefont{and}
  \bibinfo{author}{\bibfnamefont{D.-H.} \bibnamefont{Xu}},
  \bibinfo{journal}{Phys. Rev. B} \textbf{\bibinfo{volume}{97}},
  \bibinfo{pages}{155152} (\bibinfo{year}{2018}).

\bibitem[{\citenamefont{Li et~al.}(2018)\citenamefont{Li, Lee, and
  Gong}}]{Li:2018}
\bibinfo{author}{\bibfnamefont{L.}~\bibnamefont{Li}},
  \bibinfo{author}{\bibfnamefont{C.~H.} \bibnamefont{Lee}}, \bibnamefont{and}
  \bibinfo{author}{\bibfnamefont{J.}~\bibnamefont{Gong}},
  \bibinfo{journal}{Phys. Rev. Lett.} \textbf{\bibinfo{volume}{121}},
  \bibinfo{pages}{036401} (\bibinfo{year}{2018}).

\bibitem[{\citenamefont{Martin et~al.}(2017)\citenamefont{Martin, Refael, and
  Halperin}}]{Martin:2017}
\bibinfo{author}{\bibfnamefont{I.}~\bibnamefont{Martin}},
  \bibinfo{author}{\bibfnamefont{G.}~\bibnamefont{Refael}}, \bibnamefont{and}
  \bibinfo{author}{\bibfnamefont{B.}~\bibnamefont{Halperin}},
  \bibinfo{journal}{Phys. Rev. X} \textbf{\bibinfo{volume}{7}},
  \bibinfo{pages}{041008} (\bibinfo{year}{2017}).

\bibitem[{\citenamefont{Peng and Refael}(2018{\natexlab{a}})}]{Peng:2018}
\bibinfo{author}{\bibfnamefont{Y.}~\bibnamefont{Peng}} \bibnamefont{and}
  \bibinfo{author}{\bibfnamefont{G.}~\bibnamefont{Refael}},
  \bibinfo{journal}{Phys. Rev. B} \textbf{\bibinfo{volume}{98}},
  \bibinfo{pages}{220509(R)} (\bibinfo{year}{2018}{\natexlab{a}}).

\bibitem[{\citenamefont{Peng and Refael}(2018{\natexlab{b}})}]{Peng2:2018}
\bibinfo{author}{\bibfnamefont{Y.}~\bibnamefont{Peng}} \bibnamefont{and}
  \bibinfo{author}{\bibfnamefont{G.}~\bibnamefont{Refael}},
  \bibinfo{journal}{Phys. Rev. B} \textbf{\bibinfo{volume}{97}},
  \bibinfo{pages}{134303} (\bibinfo{year}{2018}{\natexlab{b}}).

\bibitem[{\citenamefont{Crowley
  et~al.}(2019{\natexlab{a}})\citenamefont{Crowley, Martin, and
  Chandran}}]{Crowley:2019}
\bibinfo{author}{\bibfnamefont{P.~J.~D.} \bibnamefont{Crowley}},
  \bibinfo{author}{\bibfnamefont{I.}~\bibnamefont{Martin}}, \bibnamefont{and}
  \bibinfo{author}{\bibfnamefont{A.}~\bibnamefont{Chandran}},
  \bibinfo{journal}{Phys. Rev. B} \textbf{\bibinfo{volume}{99}},
  \bibinfo{pages}{064306} (\bibinfo{year}{2019}{\natexlab{a}}).

\bibitem[{\citenamefont{Crowley
  et~al.}(2019{\natexlab{b}})\citenamefont{Crowley, Martin, and
  Chandran}}]{Crowley2:2019}
\bibinfo{author}{\bibfnamefont{P.~J.~D.} \bibnamefont{Crowley}},
  \bibinfo{author}{\bibfnamefont{I.}~\bibnamefont{Martin}}, \bibnamefont{and}
  \bibinfo{author}{\bibfnamefont{A.}~\bibnamefont{Chandran}},
  \bibinfo{journal}{arXiv:1908.08062}  (\bibinfo{year}{2019}{\natexlab{b}}).

\bibitem[{\citenamefont{Altland and Zirnbauer}(1997)}]{Altland:1997}
\bibinfo{author}{\bibfnamefont{A.}~\bibnamefont{Altland}} \bibnamefont{and}
  \bibinfo{author}{\bibfnamefont{M.~R.} \bibnamefont{Zirnbauer}},
  \bibinfo{journal}{Phys. Rev. B} \textbf{\bibinfo{volume}{55}},
  \bibinfo{pages}{1142} (\bibinfo{year}{1997}).

\bibitem[{\citenamefont{Schnyder et~al.}(2008)\citenamefont{Schnyder, Ryu,
  Furusaki, and Ludwig}}]{Schnyder:2008}
\bibinfo{author}{\bibfnamefont{A.~P.} \bibnamefont{Schnyder}},
  \bibinfo{author}{\bibfnamefont{S.}~\bibnamefont{Ryu}},
  \bibinfo{author}{\bibfnamefont{A.}~\bibnamefont{Furusaki}}, \bibnamefont{and}
  \bibinfo{author}{\bibfnamefont{A.~W.~W.} \bibnamefont{Ludwig}},
  \bibinfo{journal}{Phys. Rev. B} \textbf{\bibinfo{volume}{78}},
  \bibinfo{pages}{195125} (\bibinfo{year}{2008}).

\bibitem[{\citenamefont{Chiu et~al.}(2016)\citenamefont{Chiu, Teo, Schnyder,
  and Ryu}}]{ChiuReview:2016}
\bibinfo{author}{\bibfnamefont{C.-K.} \bibnamefont{Chiu}},
  \bibinfo{author}{\bibfnamefont{J.~C.~Y.} \bibnamefont{Teo}},
  \bibinfo{author}{\bibfnamefont{A.~P.} \bibnamefont{Schnyder}},
  \bibnamefont{and} \bibinfo{author}{\bibfnamefont{S.}~\bibnamefont{Ryu}},
  \bibinfo{journal}{Rev. Mod. Phys.} \textbf{\bibinfo{volume}{88}},
  \bibinfo{pages}{035055} (\bibinfo{year}{2016}).

\bibitem[{\citenamefont{Marra and Cuoco}(2017)}]{Marra:2017}
\bibinfo{author}{\bibfnamefont{P.}~\bibnamefont{Marra}} \bibnamefont{and}
  \bibinfo{author}{\bibfnamefont{M.}~\bibnamefont{Cuoco}},
  \bibinfo{journal}{Phys. Rev. B} \textbf{\bibinfo{volume}{95}},
  \bibinfo{pages}{140504(R)} (\bibinfo{year}{2017}).

\bibitem[{\citenamefont{Bauer et~al.}(2018)\citenamefont{Bauer, Karzig,
  Mishmash, Antipov, and Alicea}}]{Bauer:2018}
\bibinfo{author}{\bibfnamefont{B.}~\bibnamefont{Bauer}},
  \bibinfo{author}{\bibfnamefont{T.}~\bibnamefont{Karzig}},
  \bibinfo{author}{\bibfnamefont{R.~V.} \bibnamefont{Mishmash}},
  \bibinfo{author}{\bibfnamefont{A.~E.} \bibnamefont{Antipov}},
  \bibnamefont{and} \bibinfo{author}{\bibfnamefont{J.}~\bibnamefont{Alicea}},
  \bibinfo{journal}{SciPost Phys.} \textbf{\bibinfo{volume}{5}},
  \bibinfo{pages}{004} (\bibinfo{year}{2018}).

\bibitem[{\citenamefont{Stenger et~al.}(2018)\citenamefont{Stenger, Woods,
  Frolov, and Stanescu}}]{Stenger:2018}
\bibinfo{author}{\bibfnamefont{J.~P.~T.} \bibnamefont{Stenger}},
  \bibinfo{author}{\bibfnamefont{B.~D.} \bibnamefont{Woods}},
  \bibinfo{author}{\bibfnamefont{S.~M.} \bibnamefont{Frolov}},
  \bibnamefont{and} \bibinfo{author}{\bibfnamefont{T.~D.}
  \bibnamefont{Stanescu}}, \bibinfo{journal}{Phys. Rev. B}
  \textbf{\bibinfo{volume}{98}}, \bibinfo{pages}{085407}
  (\bibinfo{year}{2018}).

\bibitem[{\citenamefont{Haim and Stern}(2019)}]{Haim:2019}
\bibinfo{author}{\bibfnamefont{A.}~\bibnamefont{Haim}} \bibnamefont{and}
  \bibinfo{author}{\bibfnamefont{A.}~\bibnamefont{Stern}},
  \bibinfo{journal}{Phys. Rev. Lett.} \textbf{\bibinfo{volume}{122}},
  \bibinfo{pages}{126801} (\bibinfo{year}{2019}).

\bibitem[{\citenamefont{Shirley}(1965)}]{Shirley}
\bibinfo{author}{\bibfnamefont{J.~H.} \bibnamefont{Shirley}},
  \bibinfo{journal}{Phys. Rev.} \textbf{\bibinfo{volume}{138}},
  \bibinfo{pages}{979} (\bibinfo{year}{1965}).

\bibitem[{\citenamefont{D'Alessio and Rigol}(2014)}]{Dalessio:2014}
\bibinfo{author}{\bibfnamefont{L.}~\bibnamefont{D'Alessio}} \bibnamefont{and}
  \bibinfo{author}{\bibfnamefont{M.}~\bibnamefont{Rigol}},
  \bibinfo{journal}{Phys. Rev. X} \textbf{\bibinfo{volume}{4}},
  \bibinfo{pages}{041048} (\bibinfo{year}{2014}).

\bibitem[{\citenamefont{Abanin et~al.}(2015)\citenamefont{Abanin, Roeck, and
  Huveneers}}]{Abanin:2015}
\bibinfo{author}{\bibfnamefont{D.~A.} \bibnamefont{Abanin}},
  \bibinfo{author}{\bibfnamefont{W.~D.} \bibnamefont{Roeck}}, \bibnamefont{and}
  \bibinfo{author}{\bibfnamefont{F.}~\bibnamefont{Huveneers}},
  \bibinfo{journal}{Phys. Rev. Lett.} \textbf{\bibinfo{volume}{115}},
  \bibinfo{pages}{256803} (\bibinfo{year}{2015}).

\bibitem[{\citenamefont{Bukov et~al.}(2016)\citenamefont{Bukov, Heyl, Huse, and
  Polkovnikov}}]{Bukov:2016}
\bibinfo{author}{\bibfnamefont{M.}~\bibnamefont{Bukov}},
  \bibinfo{author}{\bibfnamefont{M.}~\bibnamefont{Heyl}},
  \bibinfo{author}{\bibfnamefont{D.~A.} \bibnamefont{Huse}}, \bibnamefont{and}
  \bibinfo{author}{\bibfnamefont{A.}~\bibnamefont{Polkovnikov}},
  \bibinfo{journal}{Phys. Rev. B} p. \bibinfo{pages}{155132}
  (\bibinfo{year}{2016}).

\bibitem[{\citenamefont{Mori et~al.}(2016)\citenamefont{Mori, Kuwahara, and
  Saito}}]{Mori:2016}
\bibinfo{author}{\bibfnamefont{T.}~\bibnamefont{Mori}},
  \bibinfo{author}{\bibfnamefont{T.}~\bibnamefont{Kuwahara}}, \bibnamefont{and}
  \bibinfo{author}{\bibfnamefont{K.}~\bibnamefont{Saito}},
  \bibinfo{journal}{Phys. Rev. Lett.} \textbf{\bibinfo{volume}{116}},
  \bibinfo{pages}{120401} (\bibinfo{year}{2016}).

\bibitem[{\citenamefont{Kuwahara et~al.}(2016)\citenamefont{Kuwahara, Mori, and
  Saito}}]{Kuwahara:2016}
\bibinfo{author}{\bibfnamefont{T.}~\bibnamefont{Kuwahara}},
  \bibinfo{author}{\bibfnamefont{T.}~\bibnamefont{Mori}}, \bibnamefont{and}
  \bibinfo{author}{\bibfnamefont{K.}~\bibnamefont{Saito}},
  \bibinfo{journal}{Ann. Phys.} pp. \bibinfo{pages}{96--124}
  (\bibinfo{year}{2016}).

\bibitem[{\citenamefont{Weinberg et~al.}(2017)\citenamefont{Weinberg, Bukov,
  D'Alessio, Polkovnikov, Vajna, and Kolodrubetz}}]{Weinberg:2017}
\bibinfo{author}{\bibfnamefont{P.}~\bibnamefont{Weinberg}},
  \bibinfo{author}{\bibfnamefont{M.}~\bibnamefont{Bukov}},
  \bibinfo{author}{\bibfnamefont{L.}~\bibnamefont{D'Alessio}},
  \bibinfo{author}{\bibfnamefont{A.}~\bibnamefont{Polkovnikov}},
  \bibinfo{author}{\bibfnamefont{S.}~\bibnamefont{Vajna}}, \bibnamefont{and}
  \bibinfo{author}{\bibfnamefont{M.}~\bibnamefont{Kolodrubetz}},
  \bibinfo{journal}{Phys. Rep.} \textbf{\bibinfo{volume}{688}},
  \bibinfo{pages}{1} (\bibinfo{year}{2017}).

\bibitem[{\citenamefont{Abanin et~al.}(2017{\natexlab{a}})\citenamefont{Abanin,
  Roeck, Ho, , and Huveneers}}]{Abanin:2017}
\bibinfo{author}{\bibfnamefont{D.~A.} \bibnamefont{Abanin}},
  \bibinfo{author}{\bibfnamefont{W.~D.} \bibnamefont{Roeck}},
  \bibinfo{author}{\bibfnamefont{W.~W.} \bibnamefont{Ho}}, , \bibnamefont{and}
  \bibinfo{author}{\bibfnamefont{F.}~\bibnamefont{Huveneers}},
  \bibinfo{journal}{Phys. Rev. B} \textbf{\bibinfo{volume}{95}},
  \bibinfo{pages}{014112} (\bibinfo{year}{2017}{\natexlab{a}}).

\bibitem[{\citenamefont{Abanin et~al.}(2017{\natexlab{b}})\citenamefont{Abanin,
  Roeck, Ho, and Huveneers}}]{Abanin2:2017}
\bibinfo{author}{\bibfnamefont{D.}~\bibnamefont{Abanin}},
  \bibinfo{author}{\bibfnamefont{W.~D.} \bibnamefont{Roeck}},
  \bibinfo{author}{\bibfnamefont{W.~W.} \bibnamefont{Ho}}, \bibnamefont{and}
  \bibinfo{author}{\bibfnamefont{F.}~\bibnamefont{Huveneers}},
  \bibinfo{journal}{Commun. Math. Phys.} \textbf{\bibinfo{volume}{354}},
  \bibinfo{pages}{809} (\bibinfo{year}{2017}{\natexlab{b}}).

\bibitem[{\citenamefont{Jotzu et~al.}(2014)\citenamefont{Jotzu, Messer,
  Desbuquois, Lebrat, Uehlinger, Greif, and Esslinger}}]{Jotzu:2014}
\bibinfo{author}{\bibfnamefont{G.}~\bibnamefont{Jotzu}},
  \bibinfo{author}{\bibfnamefont{M.}~\bibnamefont{Messer}},
  \bibinfo{author}{\bibfnamefont{R.}~\bibnamefont{Desbuquois}},
  \bibinfo{author}{\bibfnamefont{M.}~\bibnamefont{Lebrat}},
  \bibinfo{author}{\bibfnamefont{T.}~\bibnamefont{Uehlinger}},
  \bibinfo{author}{\bibfnamefont{D.}~\bibnamefont{Greif}}, \bibnamefont{and}
  \bibinfo{author}{\bibfnamefont{T.}~\bibnamefont{Esslinger}},
  \bibinfo{journal}{Nature} pp. \bibinfo{pages}{237--240}
  (\bibinfo{year}{2014}).

\bibitem[{\citenamefont{Messer et~al.}(2018)\citenamefont{Messer, Sandholzer,
  G\"{o}rg, Minguzzi, Desbuquois, and Esslinger}}]{Messer:2018}
\bibinfo{author}{\bibfnamefont{M.}~\bibnamefont{Messer}},
  \bibinfo{author}{\bibfnamefont{K.}~\bibnamefont{Sandholzer}},
  \bibinfo{author}{\bibfnamefont{F.}~\bibnamefont{G\"{o}rg}},
  \bibinfo{author}{\bibfnamefont{J.}~\bibnamefont{Minguzzi}},
  \bibinfo{author}{\bibfnamefont{R.}~\bibnamefont{Desbuquois}},
  \bibnamefont{and}
  \bibinfo{author}{\bibfnamefont{T.}~\bibnamefont{Esslinger}},
  \bibinfo{journal}{Phys. Rev. Lett.} \textbf{\bibinfo{volume}{121}},
  \bibinfo{pages}{233603} (\bibinfo{year}{2018}).

\bibitem[{\citenamefont{Cheng et~al.}(2019)\citenamefont{Cheng, Pan, Wang,
  Zhang, Yu, Gover, Zhang, Li, Zhou, and Zhu}}]{Cheng:2019}
\bibinfo{author}{\bibfnamefont{Q.}~\bibnamefont{Cheng}},
  \bibinfo{author}{\bibfnamefont{Y.}~\bibnamefont{Pan}},
  \bibinfo{author}{\bibfnamefont{H.}~\bibnamefont{Wang}},
  \bibinfo{author}{\bibfnamefont{C.}~\bibnamefont{Zhang}},
  \bibinfo{author}{\bibfnamefont{D.}~\bibnamefont{Yu}},
  \bibinfo{author}{\bibfnamefont{A.}~\bibnamefont{Gover}},
  \bibinfo{author}{\bibfnamefont{H.}~\bibnamefont{Zhang}},
  \bibinfo{author}{\bibfnamefont{T.}~\bibnamefont{Li}},
  \bibinfo{author}{\bibfnamefont{L.}~\bibnamefont{Zhou}}, \bibnamefont{and}
  \bibinfo{author}{\bibfnamefont{S.}~\bibnamefont{Zhu}},
  \bibinfo{journal}{Phys. Rev. Lett.} \textbf{\bibinfo{volume}{112}},
  \bibinfo{pages}{173901} (\bibinfo{year}{2019}).

\bibitem[{\citenamefont{Rubio-Abadal et~al.}(2020)\citenamefont{Rubio-Abadal,
  Ippoliti, Hollerith, Wei, Rui, Sondhi, Khemani, Gross, and
  Bloch}}]{Rubio-Abadal:2020}
\bibinfo{author}{\bibfnamefont{A.}~\bibnamefont{Rubio-Abadal}},
  \bibinfo{author}{\bibfnamefont{M.}~\bibnamefont{Ippoliti}},
  \bibinfo{author}{\bibfnamefont{S.}~\bibnamefont{Hollerith}},
  \bibinfo{author}{\bibfnamefont{D.}~\bibnamefont{Wei}},
  \bibinfo{author}{\bibfnamefont{J.}~\bibnamefont{Rui}},
  \bibinfo{author}{\bibfnamefont{S.~L.} \bibnamefont{Sondhi}},
  \bibinfo{author}{\bibfnamefont{V.}~\bibnamefont{Khemani}},
  \bibinfo{author}{\bibfnamefont{C.}~\bibnamefont{Gross}}, \bibnamefont{and}
  \bibinfo{author}{\bibfnamefont{I.}~\bibnamefont{Bloch}},
  \bibinfo{journal}{Phys. Rev. X} \textbf{\bibinfo{volume}{10}},
  \bibinfo{pages}{021044} (\bibinfo{year}{2020}).

\bibitem[{\citenamefont{Vajna et~al.}(2018)\citenamefont{Vajna, Klobas, Prosen,
  and Polkovnikov}}]{Vajna:2018}
\bibinfo{author}{\bibfnamefont{S.}~\bibnamefont{Vajna}},
  \bibinfo{author}{\bibfnamefont{K.}~\bibnamefont{Klobas}},
  \bibinfo{author}{\bibfnamefont{T.}~\bibnamefont{Prosen}}, \bibnamefont{and}
  \bibinfo{author}{\bibfnamefont{A.}~\bibnamefont{Polkovnikov}},
  \bibinfo{journal}{Phys. Rev. Lett.} \textbf{\bibinfo{volume}{120}},
  \bibinfo{pages}{200607} (\bibinfo{year}{2018}).

\bibitem[{\citenamefont{Bernevig and Hughes}(2013)}]{Bernevig13}
\bibinfo{author}{\bibfnamefont{B.~A.} \bibnamefont{Bernevig}} \bibnamefont{and}
  \bibinfo{author}{\bibfnamefont{T.~L.} \bibnamefont{Hughes}},
  \emph{\bibinfo{title}{Topological Insulators and Topological
  Superconductors}} (\bibinfo{publisher}{Princeton University Press},
  \bibinfo{year}{2013}), ISBN \bibinfo{isbn}{9780691151755}.

\bibitem[{\citenamefont{Prodan and Schulz-Baldes}(2016)}]{ProdanBook:2016}
\bibinfo{author}{\bibfnamefont{E.}~\bibnamefont{Prodan}} \bibnamefont{and}
  \bibinfo{author}{\bibfnamefont{H.}~\bibnamefont{Schulz-Baldes}},
  \emph{\bibinfo{title}{Bulk and Boundary Invariants for Complex Topological
  Insulators}} (\bibinfo{publisher}{Springer}, \bibinfo{address}{Cham
  (Switzerland)}, \bibinfo{year}{2016}).

\bibitem[{\citenamefont{Li et~al.}(2016)\citenamefont{Li, Yang, and
  Chen}}]{LinhuLi:2016}
\bibinfo{author}{\bibfnamefont{L.}~\bibnamefont{Li}},
  \bibinfo{author}{\bibfnamefont{C.}~\bibnamefont{Yang}}, \bibnamefont{and}
  \bibinfo{author}{\bibfnamefont{S.}~\bibnamefont{Chen}},
  \bibinfo{journal}{Eur. Phys. J. B} \textbf{\bibinfo{volume}{89}},
  \bibinfo{pages}{195} (\bibinfo{year}{2016}).

\bibitem[{\citenamefont{Mourik et~al.}(2012)\citenamefont{Mourik, Zuo, Frolov,
  Plissard, Bakkers, and Kouwenhoven}}]{Mourik:2012}
\bibinfo{author}{\bibfnamefont{V.}~\bibnamefont{Mourik}},
  \bibinfo{author}{\bibfnamefont{K.}~\bibnamefont{Zuo}},
  \bibinfo{author}{\bibfnamefont{S.~M.} \bibnamefont{Frolov}},
  \bibinfo{author}{\bibfnamefont{S.~R.} \bibnamefont{Plissard}},
  \bibinfo{author}{\bibfnamefont{E.~P. A.~M.} \bibnamefont{Bakkers}},
  \bibnamefont{and} \bibinfo{author}{\bibfnamefont{L.~P.}
  \bibnamefont{Kouwenhoven}}, \bibinfo{journal}{Science}
  \textbf{\bibinfo{volume}{336}}, \bibinfo{pages}{1003} (\bibinfo{year}{2012}).

\bibitem[{\citenamefont{Nadj-Perge et~al.}(2014)\citenamefont{Nadj-Perge,
  Drozdov, Li, Chen, Jeon, Seo, MacDonald, Bernevig, and
  Yazdani}}]{Nadj-Perge:2014}
\bibinfo{author}{\bibfnamefont{S.}~\bibnamefont{Nadj-Perge}},
  \bibinfo{author}{\bibfnamefont{I.~K.} \bibnamefont{Drozdov}},
  \bibinfo{author}{\bibfnamefont{J.}~\bibnamefont{Li}},
  \bibinfo{author}{\bibfnamefont{H.}~\bibnamefont{Chen}},
  \bibinfo{author}{\bibfnamefont{S.}~\bibnamefont{Jeon}},
  \bibinfo{author}{\bibfnamefont{J.}~\bibnamefont{Seo}},
  \bibinfo{author}{\bibfnamefont{A.~H.} \bibnamefont{MacDonald}},
  \bibinfo{author}{\bibfnamefont{B.~A.} \bibnamefont{Bernevig}},
  \bibnamefont{and} \bibinfo{author}{\bibfnamefont{A.}~\bibnamefont{Yazdani}},
  \bibinfo{journal}{Science} \textbf{\bibinfo{volume}{346}},
  \bibinfo{pages}{602} (\bibinfo{year}{2014}).

\bibitem[{\citenamefont{Sau and Sarma}(2012)}]{Sau:2012}
\bibinfo{author}{\bibfnamefont{J.~D.} \bibnamefont{Sau}} \bibnamefont{and}
  \bibinfo{author}{\bibfnamefont{S.~D.} \bibnamefont{Sarma}},
  \bibinfo{journal}{Nature Communications} \textbf{\bibinfo{volume}{3}},
  \bibinfo{pages}{964} (\bibinfo{year}{2012}),
  \urlprefix\url{https://doi.org/10.1038/ncomms1966}.

\bibitem[{\citenamefont{Bello et~al.}(2016)\citenamefont{Bello, Creffield, and
  Platero}}]{Bello:2016}
\bibinfo{author}{\bibfnamefont{M.}~\bibnamefont{Bello}},
  \bibinfo{author}{\bibfnamefont{C.~E.} \bibnamefont{Creffield}},
  \bibnamefont{and} \bibinfo{author}{\bibfnamefont{G.}~\bibnamefont{Platero}},
  \bibinfo{journal}{Sci. Rep.} \textbf{\bibinfo{volume}{6}},
  \bibinfo{pages}{22562} (\bibinfo{year}{2016}).

\bibitem[{\citenamefont{P\'{e}rez-Gonz\'{a}lez
  et~al.}(2019)\citenamefont{P\'{e}rez-Gonz\'{a}lez, Bello, Platero, and
  \'{A}lvaro G\'{o}mez-Le\'{o}n}}]{Perez-Gonzalez:2019}
\bibinfo{author}{\bibfnamefont{B.}~\bibnamefont{P\'{e}rez-Gonz\'{a}lez}},
  \bibinfo{author}{\bibfnamefont{M.}~\bibnamefont{Bello}},
  \bibinfo{author}{\bibfnamefont{G.}~\bibnamefont{Platero}}, \bibnamefont{and}
  \bibinfo{author}{\bibnamefont{\'{A}lvaro G\'{o}mez-Le\'{o}n}},
  \bibinfo{journal}{Phys Rev. Lett.} \textbf{\bibinfo{volume}{123}},
  \bibinfo{pages}{126401} (\bibinfo{year}{2019}).

\bibitem[{\citenamefont{Su et~al.}(1979)\citenamefont{Su, Schrieffer, and
  Heeger}}]{su79}
\bibinfo{author}{\bibfnamefont{W.~P.} \bibnamefont{Su}},
  \bibinfo{author}{\bibfnamefont{J.~R.} \bibnamefont{Schrieffer}},
  \bibnamefont{and} \bibinfo{author}{\bibfnamefont{A.~J.}
  \bibnamefont{Heeger}}, \bibinfo{journal}{Phys. Rev. Lett.}
  \textbf{\bibinfo{volume}{42}}, \bibinfo{pages}{1698} (\bibinfo{year}{1979}).

\bibitem[{\citenamefont{Rice and Mele}(1982)}]{Rice:1982}
\bibinfo{author}{\bibfnamefont{M.~J.} \bibnamefont{Rice}} \bibnamefont{and}
  \bibinfo{author}{\bibfnamefont{E.~J.} \bibnamefont{Mele}},
  \bibinfo{journal}{Phys. Rev. Lett.} \textbf{\bibinfo{volume}{49}},
  \bibinfo{pages}{1455} (\bibinfo{year}{1982}).

\bibitem[{\citenamefont{Asb{\'o}th et~al.}(2016)\citenamefont{Asb{\'o}th,
  Oroszl{\'a}ny, and P{\'a}lyi}}]{Asboth-book}
\bibinfo{author}{\bibfnamefont{J.~K.} \bibnamefont{Asb{\'o}th}},
  \bibinfo{author}{\bibfnamefont{L.}~\bibnamefont{Oroszl{\'a}ny}},
  \bibnamefont{and}
  \bibinfo{author}{\bibfnamefont{A.}~\bibnamefont{P{\'a}lyi}},
  \emph{\bibinfo{title}{A Short Course on Topological Insulators - Band
  Structure and Edge States in One and Two Dimensions}}, vol.
  \bibinfo{volume}{919} of \emph{\bibinfo{series}{Lecture Notes in Physics}}
  (\bibinfo{publisher}{Springer Verlag}, \bibinfo{year}{2016}).

\bibitem[{\citenamefont{Asboth et~al.}(2014)\citenamefont{Asboth, Tarasinski,
  and Delplace}}]{Asboth:2014}
\bibinfo{author}{\bibfnamefont{J.~K.} \bibnamefont{Asboth}},
  \bibinfo{author}{\bibfnamefont{B.}~\bibnamefont{Tarasinski}},
  \bibnamefont{and} \bibinfo{author}{\bibfnamefont{P.}~\bibnamefont{Delplace}},
  \bibinfo{journal}{Phys. Rev. B} \textbf{\bibinfo{volume}{90}},
  \bibinfo{pages}{125143} (\bibinfo{year}{2014}).

\bibitem[{\citenamefont{Fruchart}(2016)}]{Fruchart:2016}
\bibinfo{author}{\bibfnamefont{M.}~\bibnamefont{Fruchart}},
  \bibinfo{journal}{Phys. Rev. B} \textbf{\bibinfo{volume}{93}},
  \bibinfo{pages}{115429} (\bibinfo{year}{2016}).

\bibitem[{\citenamefont{Niklas et~al.}(2016)\citenamefont{Niklas, Benito,
  Kohler, and Platero}}]{Niklas:2016}
\bibinfo{author}{\bibfnamefont{M.}~\bibnamefont{Niklas}},
  \bibinfo{author}{\bibfnamefont{M.}~\bibnamefont{Benito}},
  \bibinfo{author}{\bibfnamefont{S.}~\bibnamefont{Kohler}}, \bibnamefont{and}
  \bibinfo{author}{\bibfnamefont{G.}~\bibnamefont{Platero}},
  \bibinfo{journal}{Nanotechnology} \textbf{\bibinfo{volume}{27}},
  \bibinfo{pages}{454002} (\bibinfo{year}{2016}).

\bibitem[{\citenamefont{Balabanov and Johannesson}(2017)}]{Balabanov:2017}
\bibinfo{author}{\bibfnamefont{O.}~\bibnamefont{Balabanov}} \bibnamefont{and}
  \bibinfo{author}{\bibfnamefont{H.}~\bibnamefont{Johannesson}},
  \bibinfo{journal}{Phys. Rev. B} \textbf{\bibinfo{volume}{96}},
  \bibinfo{pages}{035149} (\bibinfo{year}{2017}).

\bibitem[{\citenamefont{Creutz}(1999)}]{Creutz:1999}
\bibinfo{author}{\bibfnamefont{M.}~\bibnamefont{Creutz}},
  \bibinfo{journal}{Phys. Rev. Lett.} \textbf{\bibinfo{volume}{83}},
  \bibinfo{pages}{2636} (\bibinfo{year}{1999}).

\bibitem[{\citenamefont{Gholizadeh et~al.}(2018)\citenamefont{Gholizadeh,
  Yahyavi, and Hetenyi}}]{Gholizadeh:2018}
\bibinfo{author}{\bibfnamefont{S.}~\bibnamefont{Gholizadeh}},
  \bibinfo{author}{\bibfnamefont{M.}~\bibnamefont{Yahyavi}}, \bibnamefont{and}
  \bibinfo{author}{\bibfnamefont{B.}~\bibnamefont{Hetenyi}},
  \bibinfo{journal}{Europhys. Lett.} \textbf{\bibinfo{volume}{112}},
  \bibinfo{pages}{27001} (\bibinfo{year}{2018}).

\bibitem[{\citenamefont{Bermudez et~al.}(2009)\citenamefont{Bermudez,
  Patan\'{e}, Amico, and Martin-Delgado}}]{Bermudez:2009}
\bibinfo{author}{\bibfnamefont{A.}~\bibnamefont{Bermudez}},
  \bibinfo{author}{\bibfnamefont{D.}~\bibnamefont{Patan\'{e}}},
  \bibinfo{author}{\bibfnamefont{L.}~\bibnamefont{Amico}}, \bibnamefont{and}
  \bibinfo{author}{\bibfnamefont{M.~A.} \bibnamefont{Martin-Delgado}},
  \bibinfo{journal}{Phys. Rev. Lett.} \textbf{\bibinfo{volume}{102}},
  \bibinfo{pages}{135702} (\bibinfo{year}{2009}).

\bibitem[{\citenamefont{Mazza et~al.}(2012)\citenamefont{Mazza, Bermudez,
  Goldman, Rizzi, Martin-Delgado, and Lewenstein}}]{Mazza:2012}
\bibinfo{author}{\bibfnamefont{L.}~\bibnamefont{Mazza}},
  \bibinfo{author}{\bibfnamefont{A.}~\bibnamefont{Bermudez}},
  \bibinfo{author}{\bibfnamefont{N.}~\bibnamefont{Goldman}},
  \bibinfo{author}{\bibfnamefont{M.}~\bibnamefont{Rizzi}},
  \bibinfo{author}{\bibfnamefont{M.~A.} \bibnamefont{Martin-Delgado}},
  \bibnamefont{and}
  \bibinfo{author}{\bibfnamefont{M.}~\bibnamefont{Lewenstein}},
  \bibinfo{journal}{New J. Phys.} \textbf{\bibinfo{volume}{14}},
  \bibinfo{pages}{015007} (\bibinfo{year}{2012}).

\bibitem[{\citenamefont{J\"{u}nemann et~al.}(2017)\citenamefont{J\"{u}nemann,
  Piga, Ran, Lewenstein, Rizzi, and Bermudez}}]{Juenemann:2017}
\bibinfo{author}{\bibfnamefont{J.}~\bibnamefont{J\"{u}nemann}},
  \bibinfo{author}{\bibfnamefont{A.}~\bibnamefont{Piga}},
  \bibinfo{author}{\bibfnamefont{S.-J.} \bibnamefont{Ran}},
  \bibinfo{author}{\bibfnamefont{M.}~\bibnamefont{Lewenstein}},
  \bibinfo{author}{\bibfnamefont{M.}~\bibnamefont{Rizzi}}, \bibnamefont{and}
  \bibinfo{author}{\bibfnamefont{A.}~\bibnamefont{Bermudez}},
  \bibinfo{journal}{Phys. Rev. X} \textbf{\bibinfo{volume}{7}},
  \bibinfo{pages}{031057} (\bibinfo{year}{2017}).

\bibitem[{\citenamefont{Zurita et~al.}(2019)\citenamefont{Zurita, Creffield,
  and Platero}}]{Zurita:2019}
\bibinfo{author}{\bibfnamefont{J.}~\bibnamefont{Zurita}},
  \bibinfo{author}{\bibfnamefont{C.~E.} \bibnamefont{Creffield}},
  \bibnamefont{and} \bibinfo{author}{\bibfnamefont{G.}~\bibnamefont{Platero}},
  \bibinfo{journal}{Adv. Quantum Technol.} p. \bibinfo{pages}{1900105}
  (\bibinfo{year}{2019}).

\bibitem[{\citenamefont{Molignini
  et~al.}(2019{\natexlab{a}})\citenamefont{Molignini, Chen, and
  Chitra}}]{Molignini:2020}
\bibinfo{author}{\bibfnamefont{P.}~\bibnamefont{Molignini}},
  \bibinfo{author}{\bibfnamefont{W.}~\bibnamefont{Chen}}, \bibnamefont{and}
  \bibinfo{author}{\bibfnamefont{R.}~\bibnamefont{Chitra}},
  \bibinfo{journal}{Phys. Rev. B} \textbf{\bibinfo{volume}{101}},
  \bibinfo{pages}{165106} (\bibinfo{year}{2019}{\natexlab{a}}).

\bibitem[{\citenamefont{Chen}(2016)}]{Chen:2016}
\bibinfo{author}{\bibfnamefont{W.}~\bibnamefont{Chen}}, \bibinfo{journal}{J.
  Phys.: Condens. Matter} \textbf{\bibinfo{volume}{28}},
  \bibinfo{pages}{055601} (\bibinfo{year}{2016}).

\bibitem[{\citenamefont{Chen et~al.}(2016)\citenamefont{Chen, Sigrist, and
  Schnyder}}]{Chen-Sigrist:2016}
\bibinfo{author}{\bibfnamefont{W.}~\bibnamefont{Chen}},
  \bibinfo{author}{\bibfnamefont{M.}~\bibnamefont{Sigrist}}, \bibnamefont{and}
  \bibinfo{author}{\bibfnamefont{A.~P.} \bibnamefont{Schnyder}},
  \bibinfo{journal}{J. Phys.: Condens. Matter} \textbf{\bibinfo{volume}{28}},
  \bibinfo{pages}{365501} (\bibinfo{year}{2016}).

\bibitem[{\citenamefont{Chen et~al.}(2017)\citenamefont{Chen, Legner,
  R\"{u}egg, and Sigrist}}]{Chen:2017}
\bibinfo{author}{\bibfnamefont{W.}~\bibnamefont{Chen}},
  \bibinfo{author}{\bibfnamefont{M.}~\bibnamefont{Legner}},
  \bibinfo{author}{\bibfnamefont{A.}~\bibnamefont{R\"{u}egg}},
  \bibnamefont{and} \bibinfo{author}{\bibfnamefont{M.}~\bibnamefont{Sigrist}},
  \bibinfo{journal}{Phys. Rev. B} \textbf{\bibinfo{volume}{95}},
  \bibinfo{pages}{075116} (\bibinfo{year}{2017}).

\bibitem[{\citenamefont{Chen}(2018)}]{Chen:2018}
\bibinfo{author}{\bibfnamefont{W.}~\bibnamefont{Chen}}, \bibinfo{journal}{Phys.
  Rev. B} \textbf{\bibinfo{volume}{97}}, \bibinfo{pages}{115130}
  (\bibinfo{year}{2018}).

\bibitem[{\citenamefont{Chen and Schnyder}(2019)}]{Chen-Schnyder:2019}
\bibinfo{author}{\bibfnamefont{W.}~\bibnamefont{Chen}} \bibnamefont{and}
  \bibinfo{author}{\bibfnamefont{A.~P.} \bibnamefont{Schnyder}},
  \bibinfo{journal}{New J. Phys.} \textbf{\bibinfo{volume}{21}},
  \bibinfo{pages}{073003} (\bibinfo{year}{2019}).

\bibitem[{\citenamefont{Chen and Sigrist}(2019)}]{Chen-Sigrist-book:2019}
\bibinfo{author}{\bibfnamefont{W.}~\bibnamefont{Chen}} \bibnamefont{and}
  \bibinfo{author}{\bibfnamefont{M.}~\bibnamefont{Sigrist}},
  \emph{\bibinfo{title}{Topological Phase Transitions: Criticality,
  Universality, and Renormalization Group Approach}}
  (\bibinfo{publisher}{Wiley-Scrivener}, \bibinfo{year}{2019}).

\bibitem[{\citenamefont{Molignini
  et~al.}(2019{\natexlab{b}})\citenamefont{Molignini, Chen, and
  Chitra}}]{MoligniniReview:2019}
\bibinfo{author}{\bibfnamefont{P.}~\bibnamefont{Molignini}},
  \bibinfo{author}{\bibfnamefont{W.}~\bibnamefont{Chen}}, \bibnamefont{and}
  \bibinfo{author}{\bibfnamefont{R.}~\bibnamefont{Chitra}},
  \bibinfo{journal}{Europhys. Lett.} \textbf{\bibinfo{volume}{128}},
  \bibinfo{pages}{36001} (\bibinfo{year}{2019}{\natexlab{b}}).

\end{thebibliography}




\end{document}